\documentclass[twocolumn,superscriptaddress,amsfont,amssymb,amsmath,showpacs,balancelastpage,nofootinbib,aps]{revtex4-2}
\usepackage[utf8]{inputenc}
\usepackage{CJKutf8}

\usepackage{natbib}

\usepackage{amsmath,amssymb,amsfonts,tabu}
\usepackage{physics}
\usepackage{dsfont}
\usepackage{graphicx}
\usepackage{comment}

\usepackage[hidelinks,draft=false,colorlinks=true]{hyperref}
\usepackage[capitalize]{cleveref}

\usepackage{tikz}
\usepackage{subcaption}

\definecolor{cof}{RGB}{219,144,71}
\definecolor{pur}{RGB}{186,146,162}
\definecolor{greeo}{RGB}{91,173,69}
\definecolor{greet}{RGB}{52,111,72}

\usetikzlibrary{arrows,shapes,trees} 
\usetikzlibrary{positioning,calc}
\usetikzlibrary{decorations.markings}

\newcommand{\ra}{\rightarrow}

\newcommand{\CC}{{\mathbb C}}
\newcommand{\ZZ}{{\mathbb Z}}
\newcommand{\RR}{{\mathbb R}}

\newcommand{\cH}{{\mathcal H}}

\newcommand{\cD}{\mathcal D}

\newcommand{\cL}{\mathcal L}

\newcommand{\cO}{\mathcal O}
\newcommand{\cZ}{\mathcal Z}
\newcommand{\cF}{\mathcal F}

\newcommand{\Z}{{\mathbb Z}}
\def\U{\mathrm{U}}

\usepackage{cancel}

\usepackage[left=1in,right=1in,top=1in,bottom=1in]{geometry}

\usepackage{hyperref}

\usepackage{tikz-cd}

\hypersetup{colorlinks=true,urlcolor=[rgb]{0,0,0.5},citecolor=[rgb]{0,0.5,0},linkcolor=[rgb]{0,0.5,0}}

\makeatletter
\def\l@subsubsection#1#2{}
\makeatother

\begin{document}

\title{Detection of 2D SPT Order with Partial Symmetries}
\author{\large Alex Turzillo}
\thanks{These authors contributed equally to the work.}
\affiliation{\footnotesize Perimeter Institute for Theoretical Physics, 31 Caroline St N, Waterloo, ON N2L-2Y5, Canada}
\author{\large Naren Manjunath}
\thanks{These authors contributed equally to the work.}
\affiliation{\footnotesize Perimeter Institute for Theoretical Physics, 31 Caroline St N, Waterloo, ON N2L-2Y5, Canada}
\author{\large Jos\'e Garre-Rubio}
\affiliation{\footnotesize University of Vienna, Faculty of Mathematics, Oskar-Morgenstern-Platz 1, 1090 Vienna, Austria}
\affiliation{\footnotesize Instituto de F\'isica Te\'orica, UAM/CSIC, C. Nicol\'as Cabrera 13-15, Cantoblanco, 28049 Madrid, Spain}

\begin{abstract}

A method of using partial symmetries to distinguish two dimensional symmetry protected topological (SPT) phases of on-site, unitary symmetries is proposed. This novel order parameter takes a wavefunction, such as a ground state of a lattice model, and detects its SPT invariants as expectation values of finitely supported operators, without the need for flux insertion. The construction exploits the rotational symmetry of the lattice to extract on-site SPT invariants, building upon prior work on probing crystalline SPT phases with partial rotations. The method is demonstrated by computing the order parameter analytically on group cohomology models and numerically on a family of states interpolating between the CZX state and a trivial state. Its robustness is suggested by interpreting partial symmetries as generating the topological partition functions of lens spaces.

\end{abstract}

\maketitle

\tableofcontents

\section{Introduction}

Theoretical and experimental research on quantum matter has unveiled a diverse collection of topological phases that are distinguished from disorder by robust quantum features even in the absence of symmetry breaking \cite{doi:10.1142/S0217979290000139,Chen_2010}. Simplest among these are symmetry protected topological (SPT) orders \cite{Gu_2009,PhysRevB.81.064439,Chen_2011,Chen2013,Pollmann_2012}, whose characteristic properties like fractionalized or gapless edge modes \cite{PhysRevLett.59.799}, entanglement spectrum degeneracy \cite{PhysRevB.81.064439}, and string order \cite{PhysRevB.40.4709,PhysRevB.45.304,PhysRevB.86.125441} rely on the enforcement of the symmetry. While the abstract classification of phases is highly developed (and in the case of SPT phases, largely complete), challenges remain in characterizing phases in terms of quantities that can be measured in practice. \emph{Order parameters} dictate which measurements should be performed to extract topological information from quantum systems that arise in experiment and quantum simulation as well as in classical simulation by tensor networks \cite{Pollmann_2012,Elben2020RM,RMChern_Cian21}. In these settings, they also serve as useful tools for probing phase transitions.

Order parameters are procedures for determining the topological phase to which a quantum system belongs. While phases are typically \emph{defined} as equivalence classes of gapped systems related by smooth deformations, it is in practice difficult to determine whether such deformations exist. Therefore, to distinguish phases, it is useful to identify a \emph{topological invariant} -- a quantity that is invariant under deformations -- and an associated \emph{order parameter} -- a procedure for measuring the invariant in quantum systems. Then, if the order parameters measured in two systems disagree, it can be concluded that they lie in distinct topological phases. Engineering order parameters that extract the universal physics of topological phases independent of microscopic details is a challenging problem for which no systematic approach exists. Nevertheless, order parameters for many types of phases are known. In some cases a topological invariant has more than one known order parameter that measures it, and which procedure one prefers to use in practice varies by context.

For symmetry breaking orders (of ordinary $0$-form symmetries), a set of order parameters is given by ground state expectation values of \emph{local} operators. When an operator is charged under a symmetry, its expectation value on a state vanishes as long as the state is symmetric; that is, for ground states of symmetric systems, unless the symmetry is spontaneously broken.\footnote{In the general quantum setting, the possibility of symmetric cat states means that two-point correlators of local operators are needed to detect spontaneous symmetry breaking. But typically one is interested in the states satisfying cluster decomposition, where one-point functions suffice.} The symmetric disordered phase is characterized by the vanishing of all such order parameters. A classic example of symmetry breaking order occurs in ferromagnets: when cooled below the Curie temperature, spins align in a particular direction, breaking the rotational symmetry, and allowing the net magnetization to take a nonzero value.

In contrast, topological phases without symmetry breaking cannot be locally distinguished from the trivial disordered phase. Therefore, their detection requires new types of order parameters. Free fermion systems, for example, have their invariants encoded in the topologies of their band structures \cite{Chiu2016review,Kruthoff2017bandCombinatorics,bradlyn2017topological,Po2017symmind,watanabe2018structure,khalaf2018symmetry,tang2019comprehensive,Cano_2021}. For topological phases of interacting many-body systems, it is sometimes known how to construct \emph{nonlocal} order parameters that fill this role.

The theory of nonlocal order parameters is comprehensive in one spatial dimension. The absence of intrinsic topological order in this dimension means that the only symmetry unbroken phases are SPT phases. Many of these phases can be distinguished by their \emph{string order parameters} \cite{PhysRevB.40.4709,PhysRevB.45.304,PhysRevB.86.125441}. Twisted sector charges are another probe which detects the same topological invariants as strings \cite{ShiozakiRyu,me_KTY1}. When the protecting symmetry is on-site, unitary, and abelian, either of these order parameters suffices to completely distinguish all SPT phases. To distinguish all SPT phases for more general symmetries, order parameters based on swaps, partial inversions, and partial transposes can be employed \cite{PhysRevB.86.125441,Haegeman_2012,ShiozakiRyu}.

The classification of topological phases in two dimensions is complicated by the presence of intrinsic topological order and its interplay with symmetry in symmetry enriched topological (SET) phases. Even for SPT phases, which have no intrinsic topological order, our understanding is less comprehensive than in one dimension. Two dimensional bosonic SPT phases protected by an on-site, unitary symmetry $G$ are classified by group cohomology classes $[\omega]\in\cH^3(G,U(1))$ \cite{Chen2013}, which can be interpreted as anomalies that arise at the boundary \cite{Else2014SPT} and which determine the braiding statistics of excitations in the state obtained by gauging the symmetry \cite{Kitaev_2006,Levin_2012}. The abstract data of $[\omega]$ consists not of any directly observable number but rather of a class of cocycles modulo coboundaries; nevertheless, it is captured by various topologically invariant quantities, for which proposals exist to compute. For example one can insert a symmetry flux and observe how it binds fractional charge or binds degenerate degrees of freedom that transform projectively \cite{Zaletel_2014}. Other proposals include using membrane order parameters that generalize the strings of one dimension \cite{P_rez_Garc_a_2008}, measuring charged entanglement entropy \cite{Matsuura_2016}, measuring SPT entanglement \cite{Marvian_2017}, or extracting the invariants from SPT entanglers \cite{PhysRevB.107.235104}. Beyond SPT phases, order parameters of intrinsic topological orders have also been studied \cite{Kitaev_TEE,PhysRevB.103.125104,sheffer2025extractingtopologicalspinsbulk,liu2025detectingemergent1formsymmetries}. These include procedures to measure the many-body Chern number and the chiral central charge from a single wavefunction by using partial swaps and randomized measurements \cite{dehghani2021,Elben2020RM,RMChern_Cian21} or the modular commutator \cite{Kim_2022_a,Kim_2022_b,Fan2023HallC}.

The present proposal is to probe the SPT order of two dimensional states with a different type of order parameter, which has several advantages over existing methods. Our \emph{partial symmetry order parameter} detects the SPT order associated with the on-site symmetry $G$ by taking advantage of additional crystalline symmetries of the state. On a region of space, we act on a wavefunction with a symmetry $g\in G$ together with a rotational symmetry. The collection of expectation values of these operations, taken over various $g$ and orders of rotation, allows us (in many cases) to deduce SPT invariants for the $G$ symmetry. Partial rotations have been used previously to detect the \emph{crystalline} SPT and SET invariants of two dimensional states \cite{Shapourian2017FSPT,Shiozaki2017point,PhysRevB.98.035151,zhang2023complete,manjunath2023classif,kobayashi2024FCI,Garre_Rubio_2019}; here, we use them to learn about on-site SPT invariants.

The partial symmetry order parameter has several attractive features due to it being simply an expectation value of an operator supported on a compact region of space. In contrast with flux insertion, which involves knowing a Hamiltonian, modifying it in a nonlocal manner to insert the defect, then studying the ground states of the result; the partial symmetry order parameter operates on an unmodified given wavefunction and does so with only a small number of spatially-localized, state-independent operations. This means it can be measured on wavefunctions (such as those naturally occurring in experiment or simulation), about which one knows little besides the symmetries; in particular, finding a tensor network representation is unnecessary. Moreover, the fact that the order parameter is an expectation value of an operator makes it suitable for studying how SPT invariants evolve (such as through phase transitions), as there is a Heisenberg picture where the operator undergoes a dual evolution -- an idea that has been useful for string operators \cite{me_open1}. In many of these respects, the novel order parameter we construct resembles the order parameters (strings, swaps, partial inversions) that have been enormously useful in one dimension.

These advantages come at the cost of requiring a rotation symmetry in addition to the on-site $G$ symmetry, and this rotation complicates the extraction of the SPT invariant. Each partial symmetry expectation value is a topological invariant of the SPT phase protected by the full symmetry group, including both on-site and crystalline symmetries. These invariants may be nontrivial even when the SPT invariant for the on-site symmetry is not; this occurs when the state has nontrivial crystalline SPT invariants. Remedying this issue requires identifying the combinations of partial symmetries that isolate the desired on-site SPT invariant.

Our main results are the construction of the order parameters, their evaluations on the CZX state, a perturbation of the CZX state away from the fixed-point, and group cohomology models; a prescription for isolating an on-site SPT invariant from combinations of partial symmetries; and an interpretation of this invariant as a topological partition function. These findings constitute strong evidence that partial symmetry order parameters can distinguish many two dimensional SPT orders by computing their $\ZZ_n$ topological invariants, modulo some lattice dependent divisor of $n$.

The paper proceeds as follows. \cref{sec:partialrotation} begins with the definition of the partial symmetry order parameter on a lattice of spins. The order parameter is put to its first test on the CZX state \cite{Chen_2011}, where it detects that the state lies in the nontrivial SPT phase. We then numerically evaluate the order parameter on a family of states that interpolates between the CZX state and a symmetric product state. The order parameter is then interpreted as the lens space partition function of a topological field theory. In \cref{sec:groupcohom}, we compute the order parameter on the ground states of the group cohomology models \cite{Chen2013} that realize fixed points of all SPT phase protected by on-site, unitary symmetries on the square lattice. Doing so requires care in finding an appropriate rotation operation. \cref{sec:cryst} addresses complications that arise due the crystalline symmetry. By studying an effective topological action for background gauge fields of the symmetries, we determine how to isolate the SPT invariant for the on-site symmetry from the crystalline invariants.


\section{Partial Symmetry Order}\label{sec:partialrotation}

\subsection{The order parameter}\label{sec:definition}

Consider a two-dimensional lattice of spins with an on-site symmetry $G$ that acts via a many-body operator $\hat{U}_g$ for each $g\in G$.\footnote{We will use $g$ to denote a general group element and ${\bf g}$ to denote a generator of a $\Z_n$ subgroup.} Suppose we are given a $G$-invariant state $\ket{\Psi}$ in this Hilbert space and would like to (partially) determine its $G$-SPT order, when it is defined.\footnote{Symmetric ground states of gapped, local Hamiltonians have well-defined SPT order. Whether other states, such as excited states or ground states of critical systems, have features that are measured by partial symmetry order parameters is an interesting question for future investigation.} We solve this problem by using the rotational symmetries of the lattice.

Suppose the lattice has an $M$-fold rotation symmetry around some origin of rotations (the total rotational symmetry may be larger, but it has at least $M$). The rotation symmetry acts on the many-body Hilbert space by an operator $\hat{C}_{M}$, which both permutes the local Hilbert spaces according to the rotation of the lattice and acts on each local Hilbert space, i.e. the on-site degrees of freedom transform in some representation of the symmetry.

\begin{figure}[h]
\includegraphics[width=0.45\textwidth]{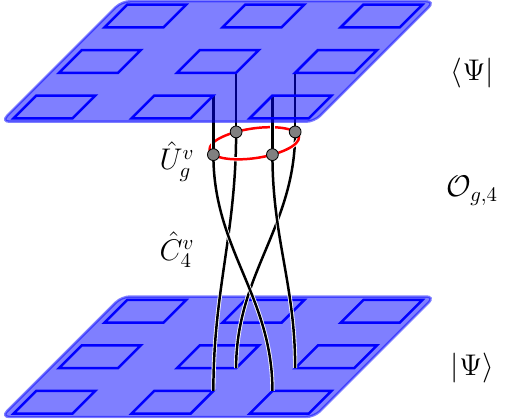}
\caption{The order parameter on a single site }
\label{fig:orderparam}
\end{figure}

Consider a region $\cD$ that is centered at the origin of rotations and is symmetric under the $M$-fold rotations. Let $\hat{U}_g^\cD$ and $\hat{C}_{M}^\cD$ be the restrictions of the symmetry operators to the lattice sites contained in $\cD$ (and acting trivially outside of $\cD$). We require that the restricted symmetries commute:\footnote{For a fixed on-site symmetry $\hat{U}_g$, this may require a clever choice of rotation $\hat{C}_{M}$. See \cref{sec:groupcohom} for an example.}
\begin{equation}
    \hat{U}_g^\cD\hat{C}_{M}^\cD=\hat{C}_{M}^\cD\hat{U}_g^\cD~.
\end{equation}
Then consider the composite operators
\begin{equation}
    \cO_{g,M}^\cD=\hat{U}_g^\cD\hat{C}_M^\cD~.
\end{equation}
We will study how the phase parts of the expectation values of these operators on a symmetric state $|\Psi\rangle$
\begin{equation}\label{orderparam}
    I_{g,M}={\rm phase}\left(\langle\Psi|\cO_{g,M}^\cD|\Psi\rangle\right)\in \text{U}(1)
\end{equation}
encode information about the SPT order. When the expectation value vanishes, the quantity $I_{g,M}$ is not defined. We will not study the modulus of the expectation value here, but the expectation is that, for states with finite correlation length $\xi$ much smaller than the perimeter $|\partial\cD|$ of $\cD$, it decays monotonically to zero in the unitless quantity $|\partial\cD|/\xi$.\footnote{A similar claim holds for fermionic Chern insulators \cite{Shiozaki2017point,zhang2023complete}.} The partial symmetry order parameter is depicted in \cref{fig:orderparam} in the case of a single site $\cD=\{v\}$.

Ultimately we are interested in the $G$-SPT order of states with a given on-site symmetry action $\hat U_g$. This order is independent of any rotational symmetries of the state; nevertheless, the partial symmetry order parameter uses a rotation operator $\hat C_M$ to extract the $G$-SPT invariants. In order for this method to make sense, its results should be independent of the choice of rotation operator. It turns out, however, that the quantities $I_{g,M}$ may depend on this choice. For example, composing the rotation with an on-site symmetry $h$ in the center of $G$ and of order dividing $M$ yields another valid rotation operator
\begin{equation}
    \hat C_M':=\hat C_M\hat U_h~.
\end{equation}
This redefinition permutes the quantities as
\begin{equation}
    I_{g,M}\,\longrightarrow\,I_{hg,M}~.
\end{equation}
Similarly, shifting the center of rotation can change the quantities $I_{g,M}$; for example, we calculate in \cref{sec:CZX} that, on the CZX state with symmetry $\ZZ_2=\{{\bf e},{\bf g}\}$, vertex centered and edge centered partial symmetries have $I^v_{{\bf g},2}=-1$ and $I^e_{{\bf g},2}=+1$, respectively. This apparent problem is resolved by understanding that the quantities $I_{g,M}$ are not simply $G$-SPT invariants, as crystalline topological invariants can complicate the story. A precise claim about how to recover a $G$-SPT invariant from a \emph{rotation-independent combination} of partial symmetry measurements will be made in \cref{sec:cryst}. For now, let us compute $I_{g,M}$ for some example states.

\subsection{Example: the CZX state}\label{sec:CZX}

As a demonstration of how the partial symmetry order parameters \eqref{orderparam} can distinguish an SPT phase from the trivial phase, consider the canonical example of a $\ZZ_2$ SPT ordered state -- the CZX state \cite{Chen_2011}. We will argue that the CZX state has the invariant $I_{{\bf g},2}=-1$, in contrast with symmetric product states which clearly have all $I=+1$.

\begin{figure}[h]
\includegraphics[width=0.45\textwidth]{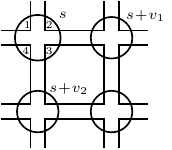}
\caption{The ground state of the CZX model.}
\label{fig:czx}
\end{figure}

The CZX state is defined on the square lattice, so $M=2$ or $4$. The Hilbert space at each site $s$ consists of $4$ subsites, each a qubit $\CC^2_{s,r}$, $r=1,2,3,4$. Work in a basis $|i_{s,r}\rangle\in\CC^2_{s,r}$ with $i_{s,r}=0,1$. The CZX state, drawn in \cref{fig:czx}, consists of a $4$-qubit GHZ state on each square plaquette of subsites:
\begin{align}\begin{split}\label{CZXstate}
    &|\Psi_{\rm CZX}\rangle=\bigotimes_s|\Psi_{\rm GHZ}[(s,3)(s+v_1,4)\\&\qquad\qquad\qquad\qquad(s+v_1+v_2,1)(s+v_2,2)]\rangle~,
\end{split}\end{align}
where $|\Psi_{\rm GHZ}[(s_1,r_1)\cdots(s_n,r_n)]\rangle$ is the GHZ state on the qubits $\CC^2_{s_i,r_i}$, and $v_{1,2}=(1,0),(0,1)$ denote the two lattice translations.

The CZX state has a rotational symmetry given by rotation of the sites as well as an on-site action that rotates the four qubits of each site in the same way. The rotation center may be taken to be a vertex, a plaquette-center, or an edge-center, in which case the rotations are $4$-fold, $4$-fold, and $2$-fold, generated by operators $\hat C_4^v$, $\hat C_4^p$, and $\hat C_2^e$, respectively.

The CZX state has an on-site $\ZZ_2$ symmetry that acts on-site by hitting each subsite with $X$ and each neighboring (non-diagonal) pair of subsites with CZ:
\begin{equation}
    U_g=CZ_{12}CZ_{23}CZ_{34}CZ_{14}X_1X_2X_3X_4~.
\end{equation}
The $X$'s, the $CZ$'s, the spatial rotations, and the intrasite rotations all commute with each other, so we can afford to be relaxed about their ordering.

The combinations of rotation and on-site symmetries give rise to ten quantities $I_{g,M}$, which evaluate as follows, where $g={\bf e},{\bf g}$ denote the elements of $\ZZ_2$:
\begin{itemize}
\item \emph{vertex centered rotations:}
\begin{equation}\label{vertexczx}
    I_{{\bf e},4}^v=1~,\quad I_{{\bf e},2}^v=1~,\quad I_{{\bf g},4}^v=1~,\quad I_{{\bf g},2}^v=-1~.
\end{equation}
\item \emph{plaquette centered rotations:}
\begin{equation}\label{plaqczx}
    I_{{\bf e},4}^p=1~,\quad I_{{\bf e},2}^p=1~,\quad I_{{\bf g},4}^p=1~,\quad I_{{\bf g},2}^p=-1~.
\end{equation}
\item \emph{edge centered rotations:}
\begin{equation}\label{edgeczx}
    I_{{\bf e},2}^e=1~,\quad I_{{\bf g},2}^e=1~.
\end{equation}
\end{itemize}
The quantities $I^{v,p,e}_{{\bf g},1}$ associated with nontrivial symmetry action ${\bf g}$ and trivial rotation ($M=1$) are ill-defined, as the expectation value of $\cO_{{\bf g},1}^\cD$ vanishes. Curiously the CZX state is distinguished from trivial states by $I_{{\bf g},2}^{v,p}=-1$ but not by the other values. In \cref{sec:cryst} we present a systematic theory of how the quantities $I_{g,M}$ depend on the SPT invariants.

Let us now compute the vertex centered quantities \eqref{vertexczx}. We leave the computation of the plaquette and edge centered quantities for \cref{app:czx}.

\begin{figure}[h]
\includegraphics[width=0.3\textwidth]{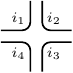}
\caption{A vertex-centered region of the CZX state.}
\label{fig:czxvertex}
\end{figure}

Consider a region $\cD$ that contains only one vertex. Write the quantity $I_{g,M}$ as sum over the basis $|i_{s,r}\rangle$ for the ket and $|i'_{s,r}\rangle$ for the bra. Contracting the parts of the GHZ loops that lie outside of $D=v$ enforces $i'_{s,r}=i_{s,r}$ for all $s,r$. The expectation value reduces to a sum over $i_{v,1},i_{v,2},i_{v,3},i_{v,4}$, which we rename $i_1,i_2,i_3,i_4$, as in \cref{fig:czxvertex}:
\begin{align}\begin{split}\label{vertexcomp}
    &\langle\Psi_{\rm CZX}|\cO_{g,M}^{(\cD=v)}|\Psi_{\rm CZX}\rangle\\
    &\sim\sum_{i_1,i_2,i_3,i_4}\langle i_1i_2i_3i_4|\hat U_g^v\hat C_M^v|i_1i_2i_3i_4\rangle\\
    &\sim\sum_{i_1,i_2,i_3,i_4}\langle i_1i_2i_3i_4|\hat U_g^v|i_{1+\frac{4}{M}}i_{2+\frac{4}{M}}i_{3+\frac{4}{M}}i_{4+\frac{4}{M}}\rangle~,\\
\end{split}\end{align}
where, in the second equality, we used that $C_M^v$ acts on $v$ by rotating the four subsites. The equalities hold up to an omitted real, positive factor.

First consider the case $g=e$ -- the pure rotation operators. With $\hat U_e^v=\mathds{1}$, the sum \eqref{vertexcomp} has no phase: it is proportional to $1$ by a real normalization factor of $1/4^M$ for $M$-fold rotations. Therefore $I_{e,M}=+1$, as claimed. Next consider the case $g={\bf g}$, where plugging $\hat U_{\bf g}^v=CZ_{12}CZ_{23}CZ_{34}CZ_{41}X_1X_2X_3X_4$ into the sum \eqref{vertexcomp} gives
\begin{equation}\label{signconstraint}
    \sum_{i_1,i_2,i_3,i_4}(-1)^{i_1i_2+i_2i_3+i_3i_4+i_4i_1}\delta_{i_k=i_{k+\frac{4}{M}}+1}~.
\end{equation}
For $M=4$, the constraint from the $\delta$ is $i_1=i_2+1=i_3=i_4+1$, so the sign becomes
\begin{equation}
    i_1(i_1+1)+(i_1+1)i_1+i_1(i_1+1)+(i_1+1)i_1=0\mod 2~,
\end{equation}
so we obtain $I_{g,4}=+1$. For $M=2$, the constraints are $i_3=i_1+1$ and $i_4=i_2+1$, so the sign becomes
\begin{equation}
    i_1i_2+i_2(i_1+1)+(i_1+1)(i_2+1)+(i_2+1)i_1=1\mod 2~,
\end{equation}
yielding a phase $I_{g,2}=-1$. For $M=1$, the constraint $i_k=i_k+1$ is not satisfied, so the sum gives $0$, which means the phase factor $I_{{\bf g},1}$ is not defined. Therefore, we have reproduced the claims in Eq.~\eqref{vertexczx}.

Conventional wisdom states that SPT orders -- unlike symmetry breaking orders -- are distinguished only by \emph{nonlocal} order parameters, so it may come as a surprise that a \emph{local} order parameter (where $\cD$ is a single vertex) was able to detect the SPT order of the CZX state. To explain this note that a region is considered nonlocal if its size is much larger than the correlation length $\xi$. (This is the notion of locality according to which a string operator must be nonlocal to distinguish SPT orders \cite{Pollmann_2012}.) Since the CZX state has correlation length $\xi=0$, the region is nonlocal no matter its size. In \cref{sec:comporderparam}, we repeat this computation on general regions $\cD$ and arrive at the same result as for a single vertex.

\subsection{Beyond the CZX fixed point}

In this section we compute the partial symmetry order parameters on states of nonzero correlation length, i.e. that are not renormalization fixed points. We study a symmetric path of states that interpolates between the CZX state and a product state of GHZ states of the four subsites of each site:
\begin{equation}
    |\rm triv\rangle=\otimes_s\left(|0000\rangle_s+|1111\rangle_s\right)~.
\end{equation}
The path goes through the symmetry broken state $\ket{0}^{\otimes N}+\ket{1}^{\otimes N}$. The perturbation we use to generate this path is similar to the one of Ref. \cite{Huang_2016} and reads
\begin{equation}
    Q(\lambda) = \mathcal{P}_{\rm GHZ} + |\lambda| \sum_{ijkl\neq 0000,1111} \ket{ijkl}\!\bra{ijkl}~,
\end{equation}
where $\mathcal{P}_{\rm GHZ} = \ket{0000}\!\bra{0000} + \ket{1111}\!\bra{1111}$. This has $Q(1)= Q(-1)= \mathds{1}$ and $Q(0)=\mathcal{P}_{\rm GHZ}$. The path is given by the following piecewise wavefunction:
\begin{equation}
\ket{\psi(\lambda)}=
    \begin{cases}
        Q(\lambda)^{\otimes N}_{\rm sites} \ket{CZX} & \text{if } \lambda \in [-1 ,0]\\
        Q(1-\lambda)^{\otimes N}_{\rm pqt} \ket{\rm triv} & \text{if } \lambda \in [0,1]
    \end{cases}
\end{equation}
which satisfies
\begin{align}\begin{split}
    \ket{\psi(-1)}&=\ket{CZX}\\
    \ket{\psi(0)}&=\ket{0}^{\otimes N}+\ket{1}^{\otimes N}\\
    \ket{\psi(1)}&=\ket{\rm triv}~,
\end{split}\end{align}
and is symmetric at all $\lambda$ since $Q(\lambda)$ commutes with the CZX symmetry on both sites and plaquettes.

\begin{figure}
\begin{center}
\includegraphics[width=7cm]{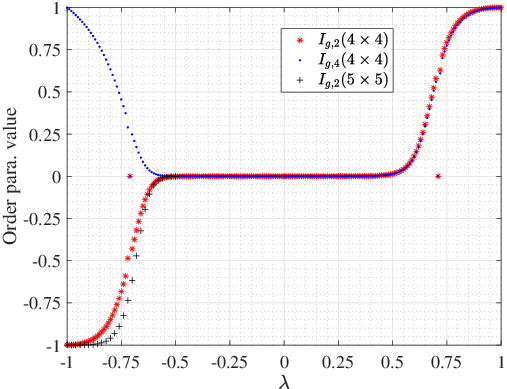}
\caption{Results of the numerical calculation with plaquette centered rotations}
\label{NumResults}
\end{center}
\end{figure}

The partial symmetry order parameters on single plaquettes are calculated exactly on tori of $4\times 4$ sites $5\times 5$ sites, with the results plotted in \cref{NumResults}. We find that $I^p_{{\bf g},2}$ varies from a plateau of $-1$ around the SPT fixed point ($\lambda=-1$) to a plateau of $+1$ around the trivial fixed point ($\lambda=+1$) via a plateau of $0$ around the global GHZ state ($\lambda=0$). The plateau of $0$ may indicate a region of spontaneous symmetry breaking bounded by phase transitions, as in Ref. \cite{Huang_2016}. The value of $I^p_{{\bf g},4}$ is $+1$ for both the SPT and the product state and also exhibits the central plateau. The width of the plateau around the SPT fixed point where $I^p_{{\bf g},2}=-1$ also seems to increase with system size. These results are in consistent with Eq.~\eqref{plaqczx}.

\subsection{Topological interpretation}\label{sec:topologicalinterp}

The universal physics of topological phases such as SPT phases is captured by topological quantum field theory (TQFT) \cite{atiyahtqft,RevModPhys.80.1083,wen2004quantum,me_KT,PhysRevB.86.125119}. One manifestation of this observation is that, in the limit of vanishing correlation length, some nonlocal order parameters of $d$ dimensional SPT phases may be interpreted as partition functions of topological field theories on certain $d+1$ dimensional spacetime topologies. Examples of order parameters with such interpretations include twisted sector charges, which capture torus partition functions \cite{ShiozakiRyu,me_KTY1,me_TY1}; partial transpose and inversion operators \cite{PhysRevB.86.125441}, which capture partition functions on the nonorientable spacetime $\RR P^2$ \cite{ShiozakiRyu,Shapourian2017FSPT}; and others \cite{Shapourian2017FSPT,PhysRevB.98.035151}. In this section, we will show that the partial symmetry order parameters $I_{g,M}$ defined in Eq.~\eqref{orderparam} may similarly be interpreted as TQFT partition functions on lens spaces.

When the SPT phase is protected by a symmetry $G$, the topological partition functions are defined on spacetimes with background $G$ gauge fields. For $G$ finite, a gauge field on $X$ is equivalent to a map $\pi_1(X)\ra G$, which assigns holonomies to cycles of $X$. The lens space $L(p;q)$, defined for $p,q$ coprime, is a $3$-manifold with fundamental group $\pi_1(L(p;q))=\ZZ_p$ generated by a single non-contractible cycle (the great circle), so a gauge field on it is parameterized by a single group element $g$ with $g^p=1$, the holonomy around the great circle. Denote the lens space with this gauge field by $L(p;q)_g$. We will argue that, on states with $p$-fold rotation symmetry, the partition function on this spacetime is captured by the partial symmetry order parameter:
\begin{equation}\label{orderislens}
    (I_{g,p})^q=\cZ(L(p;q)_g)~.
\end{equation}
When $g^p\ne 1$, the quantity $I_{g,p}$ does not have an interpretation as a partition function and is not expected to be a topological invariant of SPT phases; in \cref{sec:groupcohom} we observe how this constraint arises in lattice models. The relation \eqref{orderislens} is true for states whose crystalline SPT orders are trivial; the general relation between $I$ and $\cZ$ is discussed in \cref{sec:cryst}.

\begin{figure}
\begin{subfigure}{0.5\textwidth}
\includegraphics[width=7cm]{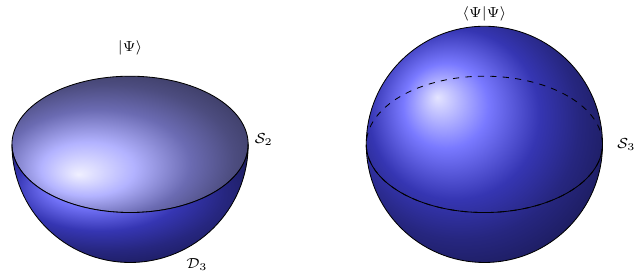}
\caption{The state and its norm on spheres in TQFT, visualized in one fewer dimension}
\label{fig:S3}
\end{subfigure}
\begin{subfigure}{0.5\textwidth}
\includegraphics[width=7cm]{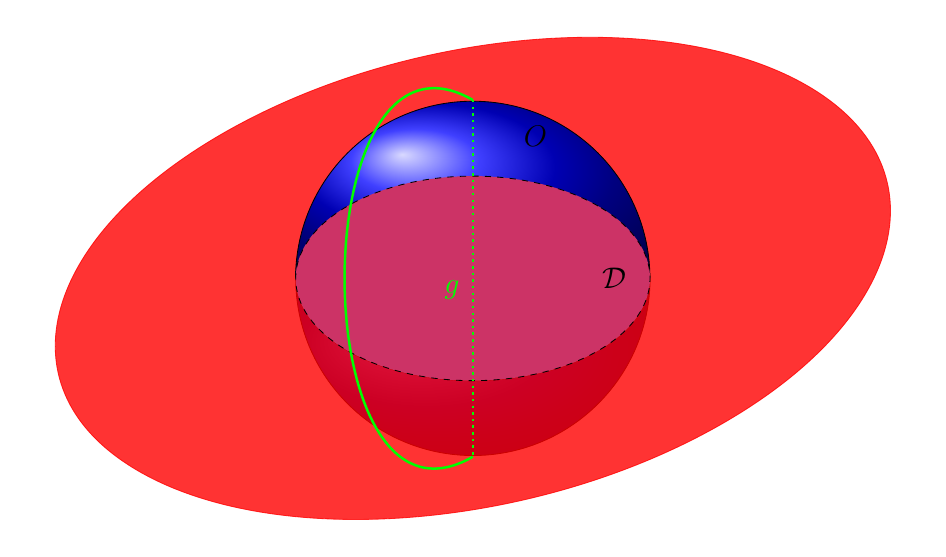}
\caption{Construction of the lens spaces. The great circle carries a holonomy $g$.}
\label{fig:lens}
\end{subfigure}
\caption{The order parameter as a lens space}
\end{figure}

First recall that the ground state wavefunction of the system on a sphere appears in TQFT as the state assigned to the solid sphere $D^3$:
\begin{equation}
    |\Psi\rangle=\cZ(D^3)\in\cH_{S^2}=\cZ(S^2)~.
\end{equation}
The norm of this state is obtained by gluing two solid spheres along their boundaries to obtain a $3$-sphere:
\begin{equation}\label{S3}
    \langle\Psi|\Psi\rangle=\cZ(S^3)=1~,
\end{equation}
as in \cref{fig:S3}. To obtain the order parameter $I_{g,M}$, the operator $\cO_{g,M}^\cD$ must be inserted into the inner product \eqref{S3}, which will change the spacetime topology from a $3$-sphere to some other $3$-manifold.

To perform this operator insertion, realize the $3$-sphere a different way, depicted in \cref{fig:lens}, as a single solid sphere with two hemispheres of its boundary identified by the identity map. The equator of $S^3$ is the spatial slice on which the state lives. A two-dimensional disk $\cD$ in this space defines a neighborhood $O$, the complement $S^3\backslash O$ of which is a solid sphere; the boundary of $S^3\backslash O$ consists of two $\cD$-shaped hemispheres, one above the equator and one below. Gluing shut the $O$-shaped ``hole'' by the identity map $D\ra D$ on its hemispheres recovers $S^3$. On the other hand, gluing it shut by a different map may result in a different closed topology. The lens space $L(p;q)$ is constructed by taking a solid sphere and identifying two hemispheres of its boundary by a rotation of angle $2\pi q/p$ \cite{hatcher}. This tells us that the expectation value of such a rotation, restricted to $\cD$, produces the partition function of $L(p;q)$.

In the operator $\cO_{g,p}^\cD$, the partial rotation by $2\pi/p$ (i.e. $q=1$) is accompanied by a partial symmetry action by $g$. This means that the holonomy of the gauge field along a path picks up a contribution of $g$ each time it crosses the equator through the region $\cD$, as in \cref{fig:lens}. The great circle represents the homotopy class of closed loops in $L(p;q)$ which cross $\cD$ exactly once; thus, the gauge field has holonomy $g$ around the great circle. We conclude that our order parameter $I_{g,p}$ is the partition function of $L(p;1)_g$.

In terms of the SPT invariant $[\omega]\in\cH^3(G,U(1))$, the lens space partition functions are given by \cite{Tantivasadakarn_2017}
\begin{equation}\label{lenspart}
    \cZ_\omega(L(p;q)_g)=\prod_{k=1}^p\omega(g,g^k,g^q)~.
\end{equation}
One can check that these quantities are gauge invariant; that is, unchanged by shifting $\omega$ by a coboundary.  Moreover, the full set of partition functions is determined by those with $q=1$, as is seen by applying the cocycle condition:
\begin{align}\begin{split}\label{Z-lenspace}
    &\cZ_\omega(L(p;q)_g)=\prod_{k=1}^p\omega(g,g^k,g^q)\\
    &\quad=\prod_{k=1}^p\frac{\omega(g^k,g,g^{q-1})}{\omega(g^{k+1},g,g^{q-1})}\omega(g,g^k,g)\omega(g,g^{k+1},g^{q-1})\\
    &\quad=\prod_{k=1}^p\omega(g,g^k,g)\omega(g,g^k,g^{q-1})\\
    &\quad=\cZ_\omega(L(p;1)_g)\cZ_\omega(L(p;q-1)_g)\\
    &\quad=\cZ_\omega(L(p;1)_g)^q~.
\end{split}\end{align}
Therefore, the general relation \eqref{orderislens} follows. Other relations satisfied by the partition functions \eqref{lenspart} are that,
for any natural number $n$,
\begin{equation}
    \cZ_\omega(L(p;1)_g)^n=\cZ_\omega(L(pn;1)_g)~,
\end{equation}
and, for any $d$ that divides $p$,
\begin{equation}
    \cZ_\omega(L(p;1)_g)^d=\cZ_\omega(L(p/d;1)_{g^d})~.
\end{equation}
These mean the order parameter satisfies
\begin{equation}\label{power}
    (I_{g,M})^n=I_{g,nM}~,\qquad\forall\, n~,\, g^M=1~,\,\,\,\quad
\end{equation}
\begin{equation}\label{Igm-secondrel}
    (I_{g,M})^d=I_{g^d,M/d}~,\quad\,\,\forall\, d\,|\,M~,\, g^M=1~.
\end{equation}
The first of these follows straightforwardly:
\begin{equation}
    \prod_{k=0}^{pn-1}\omega(g,g^k,g)    =\prod_{k=0}^{p-1}\omega(g,g^k,g)^n~,
\end{equation}
since $g^p=1$. The second relation comes from
\begin{align}\begin{split}
   &\prod_{k=0}^{p-1}\omega(g,g^k,g)^d
   =\prod_{k=0}^{p-1}\omega(g,g^k,g^d)\\
   &=\prod_{j=0}^{p/d-1}\prod_{l=0}^{d-1}\omega(g,g^l(g^d)^j,g^d)\\
   &=\prod_{j=0}^{p/d-1}\prod_{l=0}^{d-1}\frac{\omega(g^{l+1},(g^d)^j,g^d)}{\omega(g^l,(g^d)^j,g^d)} \frac{\omega(g,g^l,(g^d)^{j+1})}{\omega(g,g^l,(g^d)^j)}\\
   &=\prod_{j=0}^{p/d-1}\omega(g^d,(g^d)^j,g^d)~,
\end{split}\end{align}
where we first used the relation \eqref{Z-lenspace}, then wrote $k=jd+l$, then applied the cocycle condition for $(g,g^l,(g^d)^j,g^d)$. In the final line we noted that the second fraction vanishes under the product over $j$ and the first fraction becomes the desired result under the product over $l$.

For example, when $G=\ZZ_2$, the trivial and nontrivial SPT orders are distinguished by measuring the partition function of $L(2;1)_g=\RR P^3_g$, which is the gauge invariant quantity $\omega(g,g,g)=\pm1$. In \cref{sec:CZX}, we computed the partial symmetry order parameters of the CZX state (belonging to the nontrivial phase); for vertex- and plaquette-centered rotations, our results are consistent with their interpretation as lens space partition functions:
\begin{align}\begin{split}
    I_{{\bf e},2}&=I_{{\bf e},4}=\omega({\bf e},{\bf e},{\bf e})=+1\\
    I_{{\bf g},2}&=\omega({\bf g},{\bf g},{\bf g})=-1\\
    I_{{\bf g},4}&=(I_{{\bf g},2})^2=+1~.
\end{split}\end{align}
On the other hand, the result $I_{{\bf g},2}^e=+1$ for edge-centered rotations appears to be inconsistent with the partition function interpretation. This discrepancy suggests that the CZX state has crystalline SPT invariants for edge-centered rotations, a claim we justify in \cref{sec:cryst}.

It is known that 2D SPT orders protected by symmetry $\ZZ_n$ or $\ZZ_n\times\ZZ_m$ are completely determined by their partition functions on lens spaces $L(p;q)_g$ \cite{Tantivasadakarn_2017}. This fact suggests that partial symmetry order parameters $I_{g,M}$ form a \emph{complete set of topological invariants} for such SPT phases. More generally, SPT phases with abelian symmetry $G$ are determined by their partition functions on the lens spaces $L(p;q)_g$ and on $3$-tori $T^3_{g,h,k}$, where $g,h,k$ label holonomies around the generating cycles of $T^3$ \cite{Tantivasadakarn_2017}; therefore, in general, the partial symmetry order parameters $I_{g,M}$ only provide partial information about the SPT order. We leave the construction of an order parameter based on the $T^3$ partition function for future work.

A limitation of this topological interpretation is that, by operating in the continuum, it does not account for the crystalline symmetries of the many-body systems it purports to describe. One should worry whether the partial symmetry order parameters $I_{g,M}$ actually detect the SPT order for the internal symmetry $G$ as opposed to the crystalline or mixed internal-crystalline SPT orders. To address this subtlety, we will study the order parameters and SPT invariants more systematically in \cref{sec:cryst}.

\begin{figure}
\begin{subfigure}{0.5\textwidth}
\includegraphics[width=5.5cm]{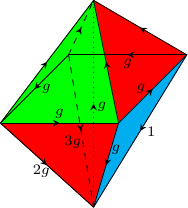}
\caption{Lens space partition function as a state sum ($p=4$)}
\label{fig:statesum}
\end{subfigure}
\begin{subfigure}{0.5\textwidth}
\includegraphics[width=7cm]{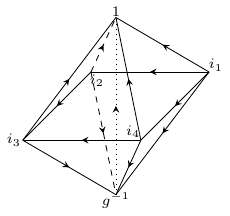}
\caption{$G$ symmetry action in the group cohomology model. An edge running between vertices labeled by $i$ and $j$ has holonomy $i^{-1}j$. The reverse orientation of two tetrahedra means their contributions enter in the denominator.}
\label{fig:groupcohom-action}
\end{subfigure}
\caption{Two triangulations of the bipyramid}
\end{figure}

The partition functions \eqref{lenspart} can be obtained from the state sum construction on the triangulation of the lens space depicted in \cref{fig:statesum} \cite{Tantivasadakarn_2017}. Gluing faces of the bipyramid that are shaded by the same color yields the lens space topology. Each of the $p$ tetrahedra contributes a factor of $\omega$, with arguments given by the edge labels of the three edges that form a chain with their arrows aligned; these contributions are precisely $\omega(g,g^k,g)$, resulting in Eq. \eqref{lenspart} in total. In \cref{sec:groupcohom}, we will encounter a different triangulation of the bipyramid, depicted in \cref{fig:groupcohom-action}. Notably the arrows in \cref{fig:groupcohom-action} are \emph{not} compatible with identifying the faces to form the lens space.

\section{Group cohomology models}\label{sec:groupcohom}

Let us now consider the group cohomology models of Chen, Gu, Liu, and Wen \cite{Chen2013}. In two dimensions, these are fixed point models that generalize the CZX model and realize, for any finite on-site symmetry $G$ and SPT invariant $\omega\in\cH^3(G,U(1))$, a state in the corresponding SPT phase. The models are symmetric under a product of the on-site $G$ symmetry and a $C_4$ lattice rotation symmetry, which we will define. After reviewing these models and their symmetries, we compute their partial symmetry order parameters on vertex- and plaquette-centered regions and discuss their relationship with the lens space partition functions, predicted in \cref{sec:topologicalinterp}. Along the way, we remark on the subtleties involved in matching the order parameter with the $G$ SPT invariant.

\subsection{Finding a rotational symmetry}

We begin by reviewing the group cohomology models \cite{Chen2013} with a focus on their realizations on the square lattice. Each site $s$ consists of $4$ subsites, each of which carries a $G$-valued quantum degree of freedom: $|i_{s,r}\rangle\in\CC_{s,r}^{|G|}$, $r=1,2,3,4$, $i_{s,r}\in G$. Like the CZX state \eqref{CZXstate}, the wavefunction is a product
\begin{equation}\label{grpcohom-state}
    |\Psi_G\rangle=\bigotimes_p\sum_{\alpha_p\in G}|\alpha_p,\alpha_p,\alpha_p,\alpha_p\rangle
\end{equation}
of GHZ-like states over square plaquettes $p$. This state has an $\omega$-dependent on-site $G$ symmetry that acts on each site $s$ as depicted in \cref{fig:groupcohom-action}:
\begin{align}\begin{split}\label{G-sym}
    U_g^s&|i_{s,1},i_{s,2},i_{s,3},i_{s,4}\rangle
    \\&\qquad=\frac{\theta_{1,2}^{s,g}\theta_{2,3}^{s,g}}{\theta_{4,3}^{s,g}\theta_{1,4}^{s,g}}|gi_{s,1},gi_{s,2},gi_{s,3},gi_{s,4}\rangle~,
\end{split}\end{align}
where $\theta_{r,r'}^{s,g}$ is shorthand for $\theta_{i_{s,r},i_{s,r'}}^g$, where
\begin{equation}
    \theta_{i,j}^g=\omega(i^{-1}j,j^{-1}g^{-1},g)~.
\end{equation}
It can be shown that this action defines a linear representation of $G$ and that the wavefunction $|\Psi_G\rangle$ is indeed invariant under it \cite{Chen2013}. Below we will assume that the symmetry $g$ used in the order parameter has $g^2$ that belongs to the center of $G$; we will point out where this assumption is needed.

This state has rotational symmetries centered on vertices, plaquettes, and edges given by rotation of the sites and rotation of the subsites within each site; however, they do not commute\footnote{On spaces like the torus, which have no boundary, the symmetries actually do commute; however, to get a meaningful order parameter, we must demand that the symmetries commute even in the presence of boundaries. If one computes the order parameter with the unmodified rotational symmetry, the result is not coboundary-invariant.} with the on-site $G$ symmetry \eqref{G-sym}, as rotations change whether contributions $\theta_{r,r'}^g$ to the phase factor appear in the numerator or denominator. Since we expect our order parameter to give meaningful results only when the symmetries commute, we will have to find a different rotational symmetry. (In the case of $G=\ZZ_2$ studied above, the phases $\theta_{r,r'}^g$ are signs, so the symmetries happen to commute without modification.)

For any choice of origin of rotations, define the rotation action on each site $s$ by
\begin{align}\begin{split}\label{rot-sym}
    &\hat C_4^s|i_{s,1},i_{s,2},i_{s,3},i_{s,4}\rangle
    \\&\,\,\,=\Lambda(i_{s,1},i_{s,2},i_{s,3},i_{s,4})|i_{s',4},i_{s',1},i_{s',2},i_{s',3}\rangle~,
\end{split}\end{align}
where $s'=C_4\cdot s$ is the site reached from $s$ by rotation and $\Lambda$ is a state-dependent phase factor. The phase factor must be chosen such that two conditions are met: the rotation must commute with $G$, and it must leave the wavefunction invariant.

The requirement that the rotation commute with the on-site symmetry on all symmetric regions $\cD$, i.e. $U_g^\cD\hat C_4^\cD=\hat C_4^\cD U_g^\cD$, imposes the constraint
\begin{align}\begin{split}
    \prod_{s\in\cD}\frac{\Lambda(gi_{s,1},gi_{s,2},gi_{s,3},gi_{s,4})}{\Lambda(i_{s,1},i_{s,2},i_{s,3},i_{s,4})}&\\&\hspace{-2cm}=\prod_{s\in\cD}\frac{\theta_{4,3}^{s,g}\theta_{1,4}^{s,g}}{\theta_{1,2}^{s,g}\theta_{2,3}^{s,g}}\frac{\theta_{4,1}^{s',g}\theta_{1,2}^{s',g}}{\theta_{3,2}^{s',g}\theta_{4,3}^{s',g}}~.
\end{split}\end{align}
By arranging the terms in the product, we can replace $s'$ with $s$. This way, we see that the condition on a single site $\cD=\{s\}$ is equivalent to the condition on arbitrary symmetric $\cD$. This condition is
\begin{equation}
    \frac{\Lambda(gi_{s,1},gi_{s,2},gi_{s,3},gi_{s,4})}{\Lambda(i_{s,1},i_{s,2},i_{s,3},i_{s,4})}=\frac{\theta_{4,1}^{s,g}\theta_{1,4}^{s,g}}{\theta_{2,3}^{s,g}\theta_{3,2}^{s,g}}~,
\end{equation}
which the bare rotation $\Lambda=1$ clearly fails. One way of satisfying this constraint is for $\Lambda$ to have the form
\begin{equation}\label{lambdaform}
    \Lambda(i_{s,1},i_{s,2},i_{s,3},i_{s,4})=\frac{\phi(i_{s,4},i_{s,1})}{\phi(i_{s,2},i_{s,3})}~,
\end{equation}
where $\phi$ are some phases that satisfy
\begin{equation}\label{g-condition}
    \frac{\phi(gi,gj)}{\phi(i,j)}=\theta_{ij}^g\theta_{ji}^g~.
\end{equation}

Meanwhile, the condition that the global action of the modified rotational symmetry leaves the wavefunction \eqref{grpcohom-state} invariant, i.e. $\langle\Psi_G|\hat C_4|\Psi_G\rangle=1$, imposes the following constraint on this phase factor:
\begin{align}\begin{split}
    &\prod_s\Lambda(i_{s,1},i_{s,2},i_{s,3},i_{s,4})=1~,\\
    &\qquad\text{for }i_{s,3}=i_{s+v_1,4}=i_{s+v_1+v_2,1}=i_{s+v_2,2}~,
\end{split}\end{align}
where $v_1,v_2$ are the lattice vectors depicted in \cref{fig:czx}. One way of satisfying this constraint is for $\Lambda$ to have the form \eqref{lambdaform}, where $\phi$ satisfies
\begin{equation}\label{f-condition}
    \frac{\phi(i,j)}{\phi(j,i)}=\frac{f(i)}{f(j)}
\end{equation}
for some phases $f$. To see why this is a solution, write the product over sites of the square lattice as a product over lattice coordinates $s=k_1v_1+k_2v_2$:
\begin{align}\begin{split}\label{f-arg}
    &\prod_s\Lambda(i_{s,1},i_{s,2},i_{s,3},i_{s,4})\\
    &\quad=\prod_s\frac{\phi(i_{s,4},i_{s,1})}{\phi(i_{s,2},i_{s,3})}=\prod_s\frac{\phi(i_{s,4},i_{s,1})}{\phi(i_{s+v_1,1},i_{s+v_1,4})}\\
    &\quad=\prod_s\frac{f(i_{s,4})}{f(i_{s,1})}=\prod_s\frac{f(i_{s+v_2,1})}{f(i_{s,1})}=1~.
\end{split}\end{align}

It remains to find a function $\phi$ that satisfies both of the conditions \eqref{g-condition} and \eqref{f-condition}. Then, via Eq. \eqref{lambdaform}, we will have an appropriate $\Lambda$. We leave the explicit construction of $\phi$ to \cref{app:moregrpcohom}, where we solve the constraints in the case of $G=\ZZ_4$. Let us remark that, because of the challenge in finding a suitable rotational symmetry, it is not obvious how to generalize this calculation to other lattices or to lattices with defects. In the remainder of this section, we will continue our computation of the partial symmetry order parameter, leaving $\phi$ written abstractly.

Before continuing, let us also mention an alternative construction. Rather than modifying the rotation symmetry to get it to commute with the on-site $G$ symmetry, we could have instead modified the $G$ symmetry and the wavefunction $|\Psi_G\rangle$. The modified $G$ action is based on the rotationally symmetric triangulation of \cref{fig:statesum}, in contrast with the usual symmetry \eqref{G-sym} which is based on the asymmetric \cref{fig:groupcohom-action}. Doing so results in a new class of models, which also depend on a cocycle $\omega$ and have the same $G$ SPT order as the usual group cohomology models. This construction is presented in \cref{app:moregrpcohom}.

\subsection{Computing the partial symmetry order}\label{sec:comporderparam}

Now we are ready to compute the partial symmetry order parameter on the group cohomology wavefunction $|\Psi_G\rangle$ \eqref{grpcohom-state}, using the $G$ symmetry \eqref{G-sym} and the rotation symmetry \eqref{rot-sym}, satisfying the conditions \eqref{g-condition}, \eqref{f-condition} above. The results are summarized in Eq. \eqref{grpcohomresults}. We use the condition that $M$ divides the order of the on-site symmetry $g$, i.e.
\begin{equation}\label{g-order}
    g^M=1~,
\end{equation}
which was introduced in \cref{sec:topologicalinterp}. The quantity $I_{g,M}$ is only defined if this condition is met; otherwise, the expectation value of $\cO_{g,M}^\cD$ vanishes. To see this, note that outside of $\cD$, the configurations $|\alpha\rangle$ and $\langle\alpha'|$ of $|\Psi_G\rangle$ and $\langle\Psi_G|$ are identified in the expectation value: $\alpha_p'=\alpha_p$. Inside of $\cD$, they are related by $\alpha_p' = g\cdot \alpha_{C_M\cdot p}$. Therefore, along the boundary $\partial\cD$, we have $\alpha_p = g\cdot \alpha_{C_M\cdot p}$, which is only possible if the condition \eqref{g-order} holds. 

We begin by computing the order parameter at a single vertex $D=\{s\}$:
\begin{align}\begin{split}
    &I_{g,4}^v
    =\langle\Psi_G|U_g^sC_4^s|\Psi_G\rangle\\
    &=\sum_{i_1,i_2,i_3,i_4}\langle i_1,i_2,i_3,i_4|U_g^sC_4^s|i_1,i_2,i_3,i_4\rangle\\
    &=\sum_{i_1,i_2,i_3,i_4}\frac{\phi(i_4,i_1)}{\phi(i_2,i_3)}\langle i_1,i_2,i_3,i_4|U_g^s|i_4,i_1,i_2,i_3\rangle\\
    &=\sum_{i_1,i_2,i_3,i_4}\frac{\phi(i_4,i_1)}{\phi(i_2,i_3)}\frac{\theta_{4,1}^g\theta_{1,2}^g}{\theta_{3,2}^g\theta_{4,3}^g}\prod_{k=1}^4\delta(g\cdot i_k=i_{[k+1]})\\
    &=\sum_{i_1,i_2,i_3,i_4}\theta_{1,2}^g\theta_{2,3}^g\theta_{3,4}^g\theta_{4,1}^g\prod_{k=1}^4\delta(g\cdot i_k=i_{[k+1]})\\
    &\qquad\times\,\cancel{\frac{\phi(i_4,i_1)}{\phi(i_2,i_3)}\frac{\phi(i_2,i_3)}{\phi(gi_2,gi_3)}\frac{\phi(i_3,i_4)}{\phi(gi_3,gi_4)}}\\
    &=\sum_{i_1}\prod_{j=1}^4\omega(i_1^{-1}gi_1,i_1^{-1}g^j,g)\\
    &=|G|\,\times\,\prod_{j=1}^4\omega(g,g^j,g)\\
    &\sim\cZ_\omega(L(4;1)_g)~.
\end{split}\end{align}
To reach the penultimate line, we used the relation
\begin{equation}\label{cocycle-product}
    \prod_j\omega(h^{-1}k,k^{-1}g^j,g)=\prod_j\frac{\omega(h^{-1},g^j,g)}{\omega(k^{-1},g^j,g)}
\end{equation}
with $h=g^{-1}i_1$ and $k=i_1$, which comes from applying the cocycle condition for $\omega$ and using the product over $j$ to cancel terms. Finally, we used the partition function \eqref{lenspart} and neglected the modulus $|G|$.

\begin{figure}
\includegraphics[width=7cm]{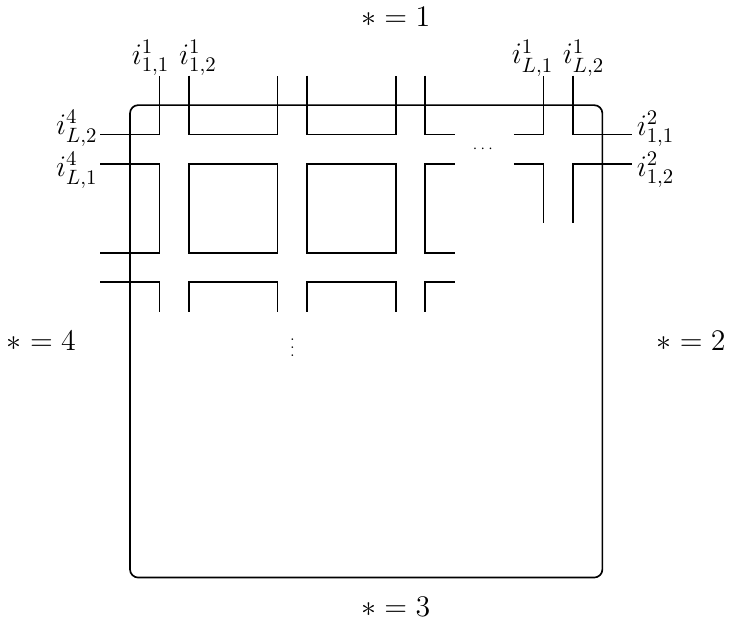}
\caption{Labels along the sides of the square}
\label{fig:edgelabels}
\end{figure}

Now let us extend the computation to square regions $\cD$ of any size, centered either on a vertex or on a plaquette. We will further generalize it to arbitrary contractible and rotationally symmetric regions in \cref{app:moregrpcohom}. The boundary $\partial\cD$ consists of four sides, each mapped into another by rotation. Denote the edges of the lattice that cross each of the four sides by $e_\ell^\star$, $\star=1,2,3,4$, $\ell=1,\ldots,L$, labeled running clockwise, as in \cref{fig:edgelabels}. Each edge $e_\ell^\star$ is associated with two subsites in $\cD$ carrying degrees of freedom $i_{\ell,1}^\star,i_{\ell,2}^\star\in |G|$. A given subsite appears in one edge or, at a corner, in two edges.

Let $\alpha$ denote a configuration of labels $i_{s,r}$ satisfying the constraint $i_{s,3}=i_{s+v_1,4}=i_{s+v_1+v_2,1}=i_{s+v_2,2}$ that each plaquette has a GHZ state, and let $\alpha'$ be the configuration resulting from the action of the symmetry. Because of this constraint, both symmetries act on the wavefunction like operators on $\partial\cD$. Let us see this for the $G$ action:
\begin{align}\begin{split}
    U_g^\cD|\Psi_G\rangle
    &=\sum_\alpha\prod_{s\in D}\frac{\theta_{1,2}^{s,g}\theta_{2,3}^{s,g}}{\theta_{4,3}^{s,g}\theta_{1,4}^{s,g}}|\alpha'\rangle\\
    &=\sum_\alpha\prod_{\ell=1}^L\frac{\theta_{i_{\ell,1}^1,i_{\ell,2}^1}^g\theta_{i_{\ell,1}^2,i_{\ell,2}^2}^g}{\theta_{i_{\ell,2}^3,i_{\ell,1}^3}^g\theta_{i_{\ell,2}^4,i_{\ell,1}^4}^g}|\alpha'\rangle~,
\end{split}\end{align}
which means each edge $e_\ell^\star$ in $\partial\cD$ contributes $\theta^g$ (if on side $1$ or $2$) or $(\theta^g)^{-1}$ (if on side $3$ or $4$). Meanwhile, by an argument like Eq. \eqref{f-arg}, the rotation action also reduces to an action on $\partial\cD$:
\begin{align}\begin{split}
    C_4^\cD&|\Psi_G\rangle
    =\sum_\alpha\prod_{s\in D}\frac{\phi(i_{s,4},i_{s,1})}{\phi(i_{s,2},i_{s,3})}|\alpha'\rangle\\
    &=\sum_\alpha\prod_{\ell=1}^L\frac{\phi(i_{\ell,1}^4,i_{\ell,2}^4)}{\phi(i_{\ell,1}^2,i_{\ell,2}^2)}\prod_{\ell=1}^{L-1}\frac{f(i_{\ell,2}^3)}{f(i_{\ell,2}^1)}~,
\end{split}\end{align}
which means each edge $e_\ell^\star$ contributes $\phi$ (if on side $4$), $\phi^{-1}$ (if on side $2$), $f$ (if on side $3$ and $\ell\ne L$), or $f^{-1}$ (if on side $1$ and $\ell\ne L$). Let us denote the factors involving $f$ as $\cF$, i.e. we define
\begin{equation*}
    \cF(\alpha):=\prod_{\ell=1}^{L-1}\frac{f(i_{\ell,2}^3)}{f(i_{\ell,2}^1)}~.
\end{equation*}

Now put these symmetry actions together:
\begin{align}\begin{split}\label{eq:grpcohomcomp}
    &I_{g,4}=\langle\Psi_G|U_g^\cD C_4^\cD|\Psi_G\rangle\\
    &=\sum_\alpha\langle\alpha|\alpha'\rangle\cF(\alpha)\prod_{\ell=1}^L\frac{\theta_{i_{\ell,1}^1,i_{\ell,2}^1}^g\theta_{i_{\ell,1}^2,i_{\ell,2}^2}^g}{\theta_{i_{\ell,2}^3,i_{\ell,1}^3}^g\theta_{i_{\ell,2}^4,i_{\ell,1}^4}^g}\frac{\phi(i_{\ell,1}^4,i_{\ell,2}^4)}{\phi(i_{\ell,1}^2,i_{\ell,2}^2)}\\
    &=\sum_\alpha\cF(\alpha)\prod_{\ell=1}^L\delta(g\cdot i_{\ell,k}^\star=i_{\ell,k}^{\star+1})\\
    &\qquad\times\,\theta_{i_{\ell,1}^1,i_{\ell,2}^1}^g\theta_{i_{\ell,1}^2,i_{\ell,2}^2}^g\theta_{i_{\ell,1}^3,i_{\ell,2}^3}^g\theta_{i_{\ell,1}^4,i_{\ell,2}^4}^g\\
    &\qquad\times\,\cancel{\frac{\phi(i_{\ell,1}^4,i_{\ell,2}^4)}{\phi(i_{\ell,1}^2,i_{\ell,2}^2)}\frac{\phi(i_{\ell,1}^3,i_{\ell,2}^3)}{\phi(gi_{\ell,1}^3,gi_{\ell,2}^3)}\frac{\phi(i_{\ell,1}^2,i_{\ell,2}^2)}{\phi(gi_{\ell,1}^2,gi_{\ell,2}^2)}}\\
    &=\pm\sum_{i_{\ell,1},i_{\ell_2}}\prod_{\ell=1}^L\prod_{j=1}^4\theta_{g^j\cdot i_{\ell,1},g^j\cdot i_{\ell,2}}^g\\
    &\qquad\times\,\delta(i_{\ell,2}=i_{\ell+1,1})\delta(i_{L,2}=g\cdot i_{1,1})\\
    &=\pm\sum_{i_{\ell,1}}\prod_{j=1}^4\omega(i_{L,1}^{-1}gi_{1,1},i_{1,1}^{-1}g^j,g)\\
    &\qquad\times\,\prod_{\ell=1}^{L-1}\omega(i_{\ell,1}^{-1}i_{\ell+1,1},i_{\ell+1,1}^{-1}g^j,g)\\
    &=\pm\sum_{i_{\ell,1}}\prod_{j=1}^4\omega(g,g^j,g)\cancel{\frac{\omega(i_{L,1}^{-1},g^j,g)}{\omega(i_{1,1}^{-1},g^j,g)}\prod_{\ell=1}^{L-1}\frac{\omega(i_{\ell,1}^{-1},g^j,g)}{\omega(i_{\ell+1,1}^{-1},g^j,g)}}\\
    &\sim\pm\cZ_\omega(L(4;1)_g)~.
\end{split}\end{align}
Here we used the fact that $\langle \alpha|\alpha'\rangle$ imposes the constraint $g\cdot i_{\ell,k}^\star=i_{\ell,k}^{\star+1}$ in order to translate the computation to one of the four sides, e.g. $\star=1$ (thus we drop the superscript on $i$). In the penultimate line, we used the cocycle relation \eqref{cocycle-product} with $h=g^{-1}i_{L,1}$, $k=i_{1,1}$ and with $h=i_{\ell,1}$, $k=i_{\ell+1,1}$. The $\pm$ sign comes from the $\cF(\alpha)$ factor. Under the constraints on the variables $i_{l,k}$, it becomes
\begin{align}\begin{split}
    \prod_{\ell=1}^{L-1}\frac{f(i_{\ell,2}^3)}{f(i_{\ell,2}^1)}&=\prod_{\ell=1}^L\frac{f(g^2\cdot i_{\ell,2}^1)}{f(i_{\ell,2}^1)}\\
    &=\prod_{\ell=1}^{L-1}\frac{\phi(g^2\cdot i_{\ell,2}^1,i_{\ell,2}^1)}{\phi(i_{\ell,2}^1,g^2\cdot i_{\ell,2}^1)}\\
    &=\prod_{\ell=1}^{L-1}\theta_{i_{\ell,2}^1,g^2\cdot i_{\ell,2}^1}^{g^2}\theta_{g^2\cdot i_{\ell,2}^1,i_{\ell,2}^1}^{g^2}\\
    &=\prod_{\ell=1}^{L-1}\frac{\omega(g^2,g^2,(i_{\ell,2}^1)^{-1})}{\omega(g^2,g^2,(i_{\ell,2}^1)^{-1}g^2)}\\
    &=\prod_{\ell=1}^{L-1}\frac{\omega(g^2,g^2,(i_{\ell,2}^1)^{-1})}{\omega(g^2,g^2,g^2(i_{\ell,2}^1)^{-1})}\\
    &=\omega(g^2,g^2,g^2)^{L-1}\\
    &=\pm 1~.
\end{split}\end{align}
Here, we used the $g^2$ symmetry to translate all terms to side $1$, then used Eq. \eqref{f-condition} followed by Eq. \eqref{g-condition} and the relation
\begin{equation}\label{thetatheta-ratio}
\theta_{\alpha,\beta}^g\theta_{\beta,\alpha}^g=\frac{\omega(\alpha^{-1}\beta,\beta^{-1}\alpha,\alpha^{-1})}{\omega(\alpha^{-1}\beta,\beta^{-1}\alpha,\alpha^{-1}g^{-1})}~,
\end{equation}
which comes from the cocycle condition for $\omega$ on $(\alpha^{-1}\beta,\beta^{-1}\alpha,\alpha^{-1}g^{-1},g)$. In the third to last line, we used the assumption (mentioned above) that $g^2$ is central in $G$. We then used the cocycle condition for $(g^2,g^2,g^2,(i_{\ell,2}^1)^{-1})$ and finally that $\omega(g^2,g^2,g^2)=\cZ_\omega(L(2;1)_{g^2})\in\{\pm 1\}$.

A vertex-centered square has a side length $L$ that is odd, so $\cF=+1$, and the computation \eqref{eq:grpcohomcomp} yields exactly the lens space partition function as predicted in \cref{sec:topologicalinterp}. A plaquette-centered square, on the other hand, has $L$ even, so $\cF$ could be either $+1$ or $-1$ (depending on $\omega$ and $g$), and the order parameter only matches the partition function up to this sign. This surprising conclusion will be explained in \cref{sec:cryst}: essentially, due to the possibility of crystalline SPT order, partial symmetry order parameters can only isolate the $\ZZ_4$ SPT index modulo $2$, so there is a sign ambiguity. In summary:
\begin{align}\begin{split}\label{grpcohomresults}
    I_{g,4}^v&=\cZ_\omega(L(4;1)_g)~,\\
    I_{g,4}^p&=\cZ_\omega(L(4;1)_g)\cZ_\omega(L(2;1)_{g^2})~.
\end{split}\end{align}

The order $2$ rotation order parameter $I_{g,2}$ for a symmetry with $g^2=1$ can be computed similarly. Here we perform the calculation for a single vertex; the generalization to a square region can be made like the above, but for brevity we will omit it here.
\begin{align}\begin{split}
    &I_{g,2}^v
    =\langle\Psi_G|U_g^sC_2^s|\Psi_G\rangle\\
    &=\sum_{i_1,i_2,i_3,i_4}\langle i_1,i_2,i_3,i_4|U_g^s(C_4^s)^2|i_1,i_2,i_3,i_4\rangle\\
    &=\sum_{i_1,i_2,i_3,i_4}\frac{\phi(i_4,i_1)}{\phi(i_2,i_3)}\frac{\phi(i_3,i_4)}{\phi(i_1,i_2)}\langle i_1,i_2,i_3,i_4|U_g^s|i_3,i_4,i_1,i_2\rangle\\
    &=\sum_{i_1,i_2,i_3,i_4}\frac{\phi(i_4,i_1)}{\phi(i_2,i_3)}\frac{\phi(i_3,i_4)}{\phi(i_1,i_2)}\frac{\theta_{3,4}^g\theta_{4,1}^g}{\theta_{2,1}^g\theta_{3,2}^g}\prod_{k=1}^4\delta(g\cdot i_k=i_{[k+2]})\\
    &=\sum_{i_1,i_2,i_3,i_4}\theta_{1,2}^g\theta_{2,3}^g\theta_{3,4}^g\theta_{4,1}^g\prod_{k=1}^4\delta(g\cdot i_k=i_{[k+2]})\\
    &\qquad\times\,\cancel{\frac{\phi(i_4,i_1)}{\phi(i_2,i_3)}\frac{\phi(i_3,i_4)}{\phi(i_1,i_2)}\frac{\phi(i_1,i_2)}{\phi(gi_1,gi_2)}\frac{\phi(i_2,i_3)}{\phi(gi_2,gi_3)}}\\
    &=\sum_{i_1,i_2}\prod_{j=0}^1\omega(i_1^{-1}i_2,i_2^{-1}g^j,g)\omega(i_2^{-1}gi_1,i_1^{-1}g^j,g)\\
    &=\sum_{i_1,i_2}\prod_{j=0}^1\omega(g,g^j,g)\cancel{\frac{\omega(g,i_2^{-1},g^j,g)}{\omega(g,i_1^{-1},g^j,g)}\frac{\omega(g,i_1^{-1},g^j,g)}{\omega(g,i_2^{-1},g^j,g)}}\\
    &\sim\cZ_\omega(L(2;1)_g)~.
\end{split}\end{align}
Here we used the relation \eqref{cocycle-product} with $h=g^{-1}i_2$, $k=i_1$ and with $h=i_1$, $k=i_2$ to manipulate the cocycle. In the last line, we neglected the modulus $|G|^2$.

\section{Systematic Characterization of SPT Invariants}\label{sec:cryst}

Recent work on SPT phases of crystalline symmetries has shown that partial rotations can also be used to characterize SPT invariants of the rotational symmetry (as opposed to the internal symmetries that are the main focus of this work) \cite{Shiozaki2017point,zhang2023complete,manjunath2023classif,kobayashi2024FCI}. These results raise the question of whether a nontrivial partial symmetry order parameter $I_{g,M}$ indicates internal (on-site) symmetry SPT order, crystalline SPT order, mixed internal-crystalline SPT order, or some combination of these. In this section, we will explain how to disentangle the internal and crystalline invariants by measuring multiple distinct partial symmetries. While a more thorough discussion of the crystalline invariants can be found in Refs.~\cite{zhang2022fractional,zhang2022pol,zhang2023complete,manjunath2023classif}, here we will focus on their application to recovering the internal SPT invariant.

\subsection{General theory}\label{sec:cryst-gen}

Consider 2d bosonic SPT states with a $G = \Z_n \times C_M$ symmetry. From group cohomology, the SPT classification is $\cH^3(G,\U(1)) \cong \Z_n \times \Z_{(n,M)} \times \Z_M$. The $\Z_n$ and $\Z_M$ factors classify SPT phases associated purely to the on-site and spatial symmetries, respectively. The $\Z_{(n,M)}$ factor, where $(n,M)$ is the greatest common divisor of $n$ and $M$, classifies mixed SPT phases of the spatial and on-site symmetries.

The invariants $(k_1,k_2,k_3)\in\ZZ_n\times\ZZ_{(n,M)}\times\ZZ_M$ appear in the effective response action
\begin{equation}\label{eq:type1-Action}
    \mathcal{L}[A,\omega] = \frac{k_1}{2\pi} A dA + \frac{k_2}{2\pi} A d\omega + \frac{k_3}{2\pi} \omega d\omega~, 
\end{equation}
where $A$ is a $\Z_n$ gauge field -- $A \sim A + 2\pi$ with holonomies of $A$ quantized in multiples of $2\pi/n$ -- and $\omega$ is a $\Z_M$ gauge field (not to be confused with the group cocycle $\omega$) -- $\omega \sim \omega + 2\pi$ with holonomies of $\omega$ quantized in multiples of $2\pi/M$. Strictly speaking, $\omega$ is a gauge field for an effective on-site $\Z_M$ symmetry rather than the spatial symmetry, but the above action nevertheless gives the correct classification of crystalline SPT phases of spins; this correspondence between the spatial and on-site classifications has been formalized as the crystalline equivalence principle and can be generalized to fermionic states as well \cite{Thorngren2018,Debray2021CEP,manjunath2022mzm}. The form of the effective action indicates that the pure on-site SPT invariant $k_1$ measures the $\ZZ_n$ charge bound to a $\ZZ_n$ flux. The second invariant is called the discrete shift and measures the $\ZZ_n$ charge bound to a lattice defect; it depends on the choice of origin and was discussed further for example in Refs.~\cite{zhang2022fractional,zhang2022pol}. The third invariant also depends on the choice of origin; it measures a `strong' SPT invariant of the rotation symmetry and is discussed further in Ref.~\cite{zhang2023complete}. We will omit the origin dependence of $k_2, k_3$ for ease of notation in the rest of Sec.~\ref{sec:cryst-gen} but restore it when we study examples.

Let us now evaluate the quantities $I_{g,M}$ in terms of the SPT invariants $k_1,k_2,k_3$. We claim that, for $g:=q \mod n$ satisfying $g^M = 1$, we have
\begin{align}\label{eq:type1-gen}
    I_{g,M}=\exp(\frac{2\pi i}{M}(\alpha^2k_1 + \alpha k_2 + k_3))~,
\end{align}
where $\alpha=qM/n$. The condition $g^M = 1$ restricts $\alpha$ to be an integer. Note that the expression \eqref{eq:type1-gen} satisfies the property $(I_{g,M})^d=I_{g^d,M/d}$ \eqref{Igm-secondrel} of lens space partition functions. The lens space relation $(I_{g,M})^n=I_{g,nM}$ \eqref{power}, on the other hand, is satisfied only if $k_2 = k_3 = 0$, reflecting how the interpretation of partial symmetry order parameters as lens space partition functions works only for states that are SPT only for the on-site symmetry.

We justify Eq.~\eqref{eq:type1-gen} in two steps. First we consider the $g={\bf e}$ case, where the claim is
\begin{equation}\label{eq:k3}
    I_{{\bf e},M} = {\rm phase}\left(\bra{\Psi} \hat{C}_{M}^\cD\ket{\Psi}\right) = e^{\frac{2\pi i}{M}k_3}~.
\end{equation}
Thus, a pure partial rotation gives the pure rotation SPT invariant $k_3$ mod $M$. This has been argued using conformal field theory in Ref.~\cite{zhang2023complete}, to which we refer the reader for a detailed discussion. The basic assumption in this calculation is that the $\rho_D$, the density matrix of the state in the region $\cD$, can be replaced with $\rho_{CFT}$, the density matrix of the edge CFT that lives on the boundary of $\cD$. Assuming this, the above expectation value can be rewritten as a trace over states in the CFT, and this in turn can be simplified in terms of quantities that directly depend on the TQFT in Eq.~\eqref{eq:type1-Action}. It is natural that $I_{{\bf e},M}$ only depends on $k_3$ and not $k_1$ or $k_2$, since the on-site symmetry is not involved.

Second, we use the result for $g=e$ \eqref{eq:k3} to derive the general expression \eqref{eq:type1-gen}. In the general order parameter, the $M$-fold rotation operator appears along with an additional $2\pi q/n$ `rotation' of the on-site $\Z_n$ symmetry by $U_g$. An elementary defect for this joint `rotation' carries an extra $2\pi q/n$ flux of the on-site symmetry compared to the a bare rotation defect. Therefore, measuring the order parameter $I_{g,M}$ in the theory $\cL[A,\omega]$ given by Eq. \eqref{eq:type1-Action} should be like measuring the pure rotation order parameter $I_{e,M}$ in the theory $\cL[A',\omega]$, where $A'$ is a gauge field modified from $A$ in order to have additional flux around defects of $\omega$; concretely,
\begin{equation}\label{eq:Arelabeled}
    A'=A+\frac{qM}{n}\omega~.
\end{equation}
Plugging $A'$ into the effective action gives
\begin{align}\begin{split}\label{eq:type1-Act-rel}
    &\mathcal L[A',\omega]=\frac{k_1}{2\pi} A dA + \frac{1}{2\pi}(2 \alpha k_1 + k_2) A d\omega\\&\qquad\quad+ \frac{1}{2\pi}\left(\alpha^2 k_1 + \alpha k_2 + k_3 \right)\omega d\omega~.
\end{split}\end{align}
By Eq. \eqref{eq:k3}, measuring $I_{e,M}$ returns the coefficient of the $\omega d\omega$ term, which in this case is the desired expression \eqref{eq:type1-gen} for $I_{g,M}$ of the original theory.

\subsection{Examples}

Consider the case $n=2, M=4$, corresponding to a state with $\ZZ_2$ on-site symmetry on a lattice with some chosen origin of $4$-fold rotations. According to the formula \eqref{eq:type1-gen}, the partial symmetries evaluate to
\begin{align}\begin{split}
    I_{{\bf e},4}&=e^{\frac{2\pi i}{4}k_3}~,\\
    I_{{\bf g},2}&=e^{\frac{2\pi i}{4}(2k_1+2k_2+2k_3)}~,\\
    I_{{\bf g},4}&=e^{\frac{2\pi i}{4}(2k_2+k_3)}~.
\end{split}\end{align}
Therefore, the following combinations are enough to fully characterize $k_1\in\ZZ_2$, $k_2\in\ZZ_2$ and $k_3\in\ZZ_4$:
\begin{align}\begin{split}\label{eq:k1-invt}
    e^{\frac{2\pi i}{2}k_1} &= I_{{\bf g},2}I_{{\bf g},4}I_{{\bf e},4}~,\\
    e^{\frac{2\pi i}{2}k_2} &= I_{{\bf e},4}/I_{{\bf g},4}~, \\
    e^{\frac{2\pi i}{4}k_3} &= I_{{\bf e},4}~.
\end{split}\end{align}
When applied to the CZX state with rotations about a vertex or plaquette, the invariants $I_{{\bf e},4}=I_{{\bf g},4}=1$, $I_{{\bf g},2}=-1$ computed in \cref{sec:CZX} and \cref{app:czx} mean that $k_1=1$, $k_2^v=k_2^p=0$, $k_3^v=k_3^p=0$. In other words, the CZX state has nontrivial SPT order for the on-site $\ZZ_2$ symmetry and trivial mixed and crystalline SPT order. Note that if one instead choses an origin of rotations with only $2$-fold rotations (e.g. the edge of a square lattice), it is impossible to disentangle the on-site, mixed, and crystalline SPT invariants, as one can only measure $k_3\,{\rm mod}\,2$ (from $I_{{\bf g},2}$) and $k_1+k^e_2+k^e_3\,{\rm mod}\,2$ (from $I_{{\bf g},2}$). For the CZX model at an edge, these quantities are both $0\,{\rm mod}\,2$. However, since $k_1$ is independent of the rotation center, the result $k_1=1$ can be used to conclude that $k^e_2=1$ for edge centered rotations, which is to say that the CZX state has mixed internal-crystalline SPT order in addition to its familiar on-site $\ZZ_2$ SPT order. In total, the SPT invariants of the CZX state are
\begin{itemize}
\item \emph{vertex centered rotations:}
\begin{equation}
    k_1=1~,\quad k_2^v=0~,\quad k_3^v=0~.
\end{equation}
\item \emph{plaquette centered rotations:}
\begin{equation}
    k_1=1~,\quad k_2^p=0~,\quad k_3^p=0~.
\end{equation}
\item \emph{edge centered rotations:}
\begin{equation}
    k_1=1~,\quad k_2^e=1~,\quad k_3^e=0~.
\end{equation}
\end{itemize}

On the other hand, for SPT states with $\Z_2$ on-site symmetry on the honeycomb lattice (such as in the Levin-Gu model \cite{Levin_2012}), partial symmetries of order $M=2,3,6$ are \textit{insufficient} to distinguish $k_1\in\ZZ_2$ from $k_2\in\ZZ_2$. The invariants in this case are
\begin{align}\begin{split}
    I_{{\bf e},6} &= e^{\frac{2\pi i}{6} k_3}~,\\
    I_{{\bf g},2} &= e^{\frac{2\pi i}{6}(3k_1 + 3k_2 + 3k_3)}~,\\
    I_{{\bf g},6} &= e^{\frac{2\pi i}{6}(3k_1 + 3k_2 + k_3)}~,
\end{split}\end{align}
with $I_{{\bf g},3}$ undefined since $\alpha=3/2\notin\ZZ$ in this case. The sum of on-site and mixed invariants $k_1+k_2$ is determined by the ratio $I_{{\bf g},6}/I_{{\bf e},6}$, but there is no combination that computes $k_1$ and $k_2$. Therefore, an independent method for finding $k_2$ (for example, computing the fractional $\Z_n$ charge at a disclination defect as in Ref. \cite{zhang2022fractional}) is needed to determine $k_1$.

Finally, consider the symmetries of the $\ZZ_4$ group cohomology model on the square lattice: $M = 2,4$ and $n=4$. In this case, we can measure
\begin{align}\begin{split}
    I_{{\bf e},4} &= e^{\frac{2\pi i}{4} k_3}~,\\
    I_{{\bf g},4} &= e^{\frac{2\pi i}{4}(k_1 + k_2 + k_3)}~,\\
    I_{{\bf g}^2,4} &= e^{\frac{2\pi i}{4}(2 k_2 + k_3)}~,\\
    I_{{\bf e},2}&=I_{{\bf e},4}^2~,\quad I_{{\bf g}^2,2}= I_{{\bf g},4}^2~,
\end{split}\end{align}
where ${\bf g}$ generates $\ZZ_4$. From these equations one can determine $k_1$ and $k_2$ mod $2$ and $k_1 + k_2$ mod $4$ as
\begin{align}\begin{split}
    e^{\frac{2\pi i}{4}(k_1+k_2)} &= I_{{\bf g},4}/I_{{\bf e},4}~,\\
    e^{\frac{2\pi i}{4}2k_1} &= I_{{\bf g},4}^2/I_{{\bf g}^2,4}I_{{\bf e},4}~, \\
    e^{\frac{2\pi i}{4}k_3} &= I_{{\bf g},4}
\end{split}\end{align}
but cannot determine $k_1$ absolutely. This explains the surprising findings of \cref{sec:groupcohom}. Recall that
\begin{align}\begin{split}
    I_{{\bf e},4}^v &= \cZ_\omega(L(4;1)_{\bf e})=1~,\\
    I_{{\bf g},4}^v &= \cZ_\omega(L(4;1)_{\bf g})=i~,\\
    I_{{\bf g}^2,4}^v &= \cZ_\omega(L(4;1)_{{\bf g}^2})=1~,\\
\end{split}\end{align}
on the vertex; whereas, due to the $\cF(\alpha)$ factor,
\begin{align}\begin{split}
    I_{{\bf e},4}^p &= \cZ_\omega(L(4;1)_{\bf e})\cZ_\omega(L(2;1)_{\bf e})=1~,\\
    I_{{\bf g},4}^p &= \cZ_\omega(L(4;1)_{\bf g})\cZ_\omega(L(2;1)_{{\bf g}^2})=-i~,\\
    I_{{\bf g}^2,4}^p &= \cZ_\omega(L(4;1)_{{\bf g}^2})\cZ_\omega(L(2;1)_{\bf e})=1~,\\
\end{split}\end{align}
on the plaquette. In the second equality of each line, we have evaluated the expression for a generating cocycle of $G=\ZZ_4$. Therefore, we can determine
\begin{align}\begin{split}
    k_1+k_2^v &= 1\mod 4~,\\
    k_1 &= 1\mod 2~, \\
    k_3^v &= 0\mod 4~,
\end{split}\end{align}
and
\begin{align}\begin{split}
    k_1+k_2^p &= 3\mod 4~,\\
    k_1 &= 1\mod 2~, \\
    k_3^p &= 0\mod 4~.
\end{split}\end{align}
From partial symmetries alone, we cannot determine the invariants absolutely. However, given outside knowledge that $k_1=1$ mod $4$ \cite{Chen2013}, we can use these relations to conclude that $k_2^v=0$ and $k_2^p=2$ for the particular choice of rotation we have used.

In \cref{app:determine}, we consider general $M$ and $n$ and study the extent to which the SPT invariants $k_1$, $k_2$, and $k_3$ are measured by partial symmetries.

\subsection{Robustness of the on-site SPT invariant}\label{sec:cryst-robust}

The invariant $k_1$ is a function only of the on-site symmetry $G$ SPT order, so it should not depend on the particulars of the rotational symmetry. However, recall from the discussion in \cref{sec:definition} that the partial symmetries $I_{g,M}$ do in fact depend on a choice of rotation operator. To resolve this tension, let us check that the \emph{combinations} of partial symmetries that build $k_1$ (as in Eqs. \eqref{eq:k1-invt}) are robust both to relocating the origin or rotations and to redefining the rotation operator by an on-site symmetry.

Defining the rotation gauge field $\omega$ requires choosing an origin within the real-space unit cell of the crystalline lattice, and the quantities $k_2, k_3$ will generally depend on this choice. Let us use the effective action \eqref{eq:type1-Action} to see that changing the origin does not affect the on-site SPT invariant $k_1$. A shift of origin can be implemented by replacing $\omega$ with some functional of $\omega$ and translation gauge fields $\vec{R}$ (explicit formulas are given in Ref. \cite{manjunath2023classif}). While the precise form of this functional depends on details of the space group, the crucial detail is that the relabeling does not involve the on-site gauge field $A$. As a result, the shift of origin leaves the $A dA$ term of the action invariant, implying that the on-site symmetry SPT invariant $k_1$ is independent of the choice of rotation center.

To see that $k_1$ is unchanged by a redefinition of the rotational symmetry, observe that redefining the rotation operator by an on-site symmetry corresponds to adding an on-site symmetry flux to every rotation flux. In the effective response action, this redefinition is implemented as Eq.~\eqref{eq:Arelabeled}. Then Eq. \eqref{eq:type1-Act-rel} demonstrates that the redefinition fixes the coefficient of $AdA$, which is the on-site invariant $k_1$, while changing the crystalline invariants $k_2, k_3$. We can also check that the explicit formulas for $k_1$ are unaffected by the redefinition. Consider the case of $n=2, M=4$. Replacing the pure rotation $\hat C_4$ by $U_g \hat{C}_4$ permutes the invariants according to
\begin{align}\begin{split}
    I_{{\bf e},4}& \,\longleftrightarrow\, I_{{\bf g},4}\\
    I_{{\bf g},2}&\quad\,\text{fixed}~,
\end{split}\end{align}
where the second relation holds because the order $2$ rotation is fixed: $\hat C_2\rightarrow U_{\bf g}^2\hat C_2=\hat C_2$. Then, looking at the expression \eqref{eq:k1-invt} for $k_1$, we see that it is invariant under this permutation. For the other choices of $n$ and $M$ discussed in \cref{app:determine}, we can perform a similar analysis to confirm that $k_1$ is invariant.

\section{Discussion}

In this paper, we have demonstrated a method to diagnose two dimensional SPT phases of on-site symmetries by using partial symmetry order parameters. We study two main examples: the square lattice CZX state, which has an on-site $\Z_2$ symmetry, and the square lattice group cohomology model with an on-site symmetry $G$. For the CZX state, we recover the expected on-site SPT invariant and also discover a crystalline SPT invariant which is protected jointly by the on-site symmetry and edge-centered rotations. We also show that the order parameter accurately diagnoses the SPT invariant in the group cohomology model; the main hurdle we overcome to do this is identifying a rotation operator that is compatible with the $G$ symmetry. Using analytical arguments based on TQFT, we argue that the partial symmetry order parameter measures a robust topological invariant. We provide further numerical evidence for this claim by perturbing the CZX state into a symmetry-breaking and then a trivial symmetric phase and observing that the order parameter clearly distinguishes these three phases. Since partial symmetry order parameters are sensitive to both on-site and crystalline SPT invariants, it is necessary to find a combination of them that isolates the desired on-site SPT invariant. Another result is to explicitly identify these combinations for various lattices, in cases where they exist.

A limitation of the partial symmetry order parameter is that it only detects the SPT invariants modulo some divisor of the order of lattice rotations; see \cref{app:determine} for a detailed discussion. This means, for example, that partial symmetries cannot extract a $\Z_n$ SPT invariant on a square lattice if $n$ is odd. A potential workaround is get a $\Z_n$ rotational symmetry locally by considering the original lattice with a disclination, which is a defect of the rotational symmetry. Theoretically, the invariant of a $\ZZ_4$ SPT state could be absolutely (rather than mod $2$ on the square lattice) determined using a disclination with $8$-fold rotations. However, properly defining the wavefunction at the disclination center can be subtle \cite{zhang2022fractional,zhang2022pol}. And as we saw in \cref{sec:groupcohom}, defining the rotation operator is not straightforward either.

Partial symmetries are not expected to distinguish all two dimensional SPT phases. Their connection to lens space partition functions (c.f. \cref{sec:topologicalinterp}) suggests that, for finite abelian symmetry groups and by exploiting suitable rotational symmetries, they can in principle distinguish all Type I and Type II cohomology classes (i.e. all classes for groups $\ZZ_n$ and $\ZZ_n\times\ZZ_m$), but not all Type III classes \cite{propitius1995topological,Tantivasadakarn_2017}. Fully distinguishing the latter requires knowing their partition functions on the $3$-torus, and it is an interesting direction for future research to engineer a new order parameter based on this space. In the language of flux insertion order parameters \cite{Zaletel_2014}, partial symmetries are able to detect SPT phases by the fractional charges bound to fluxes but cannot see SPT phases marked by projective representations attached to fluxes. Membrane operators, in contrast, only detect Type III phases,\footnote{A partial symmetry $U_g^\cD$ acts on an SPT state like an operator on $\partial\cD$. With respect to conjugation by $G$, this operator has a 1D SPT invariant given by the slant product $i_g\omega$ of $\omega$. (In other words, the $g$ domain wall is decorated by the 1D slant product SPT phase.) This slant product is what the membrane operator expectation value detects. Type I and II phases have trivial slant products, so membranes cannot distinguish them from the trivial phase. This is in contrast with SPT phases in one dimension, where string operators again detect slant products but this time slant products distinguish many phases (all, when $G$ is abelian).} so there is no overlap with partial symmetry order parameters.

Our analysis leaves open whether the order parameter is robust to evolving the state through suitably symmetric circuits, as the connection to topological invariants suggests. The answer to this question likely involves a more systematic understanding of SPT states away from renormalization fixed points, which means using regions of finite size in which the order parameter exhibits a scaling behavior. It would be interesting to study the order parameter on a broader class of states; for example, by performing numerical tests on states without translation invariance. Future analysis could look beyond SPT states to topologically ordered systems (where pure partial rotations have been studied previously \cite{kobayashi2024FCI}), to excited states, or to gapless systems.

\begin{center}
    \bf ACKNOWLEDGMENTS
\end{center}

N.M. thanks Maissam Barkeshli, Vladimir Calvera, Ryohei Kobayashi and Yuxuan Zhang for collaboration on related work. Research at Perimeter Institute is supported in part by the Government of Canada through the Department of Innovation, Science and Economic Development and by the Province of Ontario through the Ministry of Colleges and Universities. This research is funded in part by the European Union’s Horizon 2020 research and innovation program through Grant No. 863476 (ERC-CoG SEQUAM). J.G.R. acknowledges funding by the FWF Erwin Schr\"odinger Program (Grant DOI 10.55776/J4796).


\appendix

\section{Invariants of the CZX state}\label{app:czx}

In this appendix, we compute the edge- and placquette-centered invariants $I^e_{g,M}$ and $I^p_{g,M}$ of the CZX model ground state. The computation follows a similar logic to that of $I^v_{g,M}$ in \cref{sec:CZX}.

\begin{figure}[h]
\includegraphics[width=0.4\textwidth]{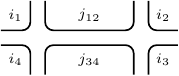}
\caption{An edge-centered region of the CZX state.}
\label{fig:czxedge}
\end{figure}

First take the origin of rotations to be the center of an edge $e$ and the region $\cD$ to consist of the two sites at the ends of $e$, as in \cref{fig:czxedge}.

Write the invariant as a sum over the indices in \cref{fig:czxedge}. As in the vertex computation, contracting the parts of the GHZ loops that lie outside the support of the order parameter enforces that the bra and ket indices are the same. And a similar argument shows that the case of pure rotation $g={\bf e}$ results in invariants $I_{{\bf e},M}=+1$.

Consider the symmetry $g={\bf g}$. Similarly to the vertex case \eqref{signconstraint}, the invariant is given by a sum over a sign and a $\delta$ constraint. Note that $j_{12}$ and $j_{34}$ are connected by two CZ gates, so their joint contribution cancels; in total, the sign is
\begin{equation}
    i_1i_4+i_2i_3+j_{12}(i_1+i_2)+j_{34}(i_3+i_4)~.
\end{equation}
For $M=2$, the constraint is $i_3=i_1+1$, $i_4=i_2+1$, $j_{34}=j_{12}+1$, so the sign becomes
\begin{align}\begin{split}
    &i_1(i_2+1)+i_2(i_1+1)+j_{12}(i_1+i_2)\\
    &\qquad\qquad+(j_{12}+1)(i_1+i_2+2)\stackrel{2}{=}0~,
\end{split}\end{align}
so the invariant is $I_{{\bf g},2}=+1$. For $M=1$, the constraint cannot be satisfied, so the sum vanishes and the invariant $I_{{\bf g},1}$ is ill-defined.

\begin{figure}[h]
\includegraphics[width=0.4\textwidth]{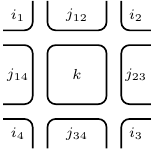}
\caption{A plaquette-centered region of the CZX state.}
\label{fig:czxplaq}
\end{figure}

Now take the origin of rotations to be the center of a plaquette $p$ and the region $\cD$ to consist of the four sites at the corners of $p$, as in \cref{fig:czxplaq}.

Write the invariant as a sum over the indices in \cref{fig:czxplaq}. Contracting the part outside the support of the order parameter enforces that the bra and ket indices are the same at each of the four corners and four sides (e.g. $i_1=i_1'$ and $j_{12}=j_{12}'$, where the prime denotes the bra index), so in total we have indices $i_1, i_2, i_3, i_4, j_{12}, j_{13}, j_{24}, j_{34}, k$, and $k'$. As usual, the case $g={\bf e}$ results in invariants $I_{e,M}=+1$.

Consider $g={\bf g}$. Similarly to Eq. \eqref{signconstraint}, the invariant is a sum over a sign and a $\delta$ constraint. Note that each $j$ is connected to $k$ by two CZ gates, so these signs will cancel; in total, the sign is
\begin{align}\begin{split}
    &j_{12}(i_1+i_2)+j_{23}(i_2+i_3)\\
    &\qquad\qquad+j_{34}(i_3+i_4)+j_{14}(i_1+i_4)~.
\end{split}\end{align}
For $M=4$, the constraints are $i_1=i_2+1=i_3=i_4+1$ and $j_{12}=j_{23}+1=j_{34}=j_{14}+1$, so the sign becomes, and so the invariant is $I_{g,4}=+1$. For $M=2$, the constraints are $i_3=i_1+1$, $i_4=i_2+1$, $j_{34}=j_{12}+1$, $j_{14}=j_{23}+1$, so the sign becomes $(-1)^1$, and so the invariant is $I_{{\bf g},2}=-1$. And finally for $M=1$, the constraint is impossible to satisfy, so the invariant $I_{{\bf g},1}$ is ill-defined.

In conclusion, we've recovered the claimed SPT invariants of the CZX state for rotations centered on an edge \eqref{edgeczx} and on a plaquette \eqref{plaqczx}.

\section{More on group cohomology models}\label{app:moregrpcohom}

\subsubsection{An explicit solution for $\phi$}

Here we write an explicit solution for $\phi$ that satisfies the conditions \eqref{g-condition} and \eqref{f-condition}. We consider the case of $G = \ZZ_4=\{0,1,2,3\}$, with the group operation written as addition. Consider
\begin{equation}\label{phi-solution}
    \phi(\alpha,\beta)=\frac{C(\alpha-\beta)}{\omega(\beta-\alpha,\alpha-\beta,-\alpha)}~,
\end{equation}
where
\begin{align}\begin{split}
    C(0)&=C(1)=C(2)=1~,\\C(3)&=f(3)\,\omega(1,3,1)
\end{split}\end{align}
and $f$ can be taken to be the partition functions \eqref{lenspart}
\begin{equation}
    f(\alpha)=\cZ_\omega(4;\alpha)_1=\omega(1,1,\alpha)\omega(1,2,\alpha)\omega(1,3,\alpha)~.
\end{equation}
Recall that these satisfy Eq.~\eqref{Z-lenspace}, which implies
\begin{equation}
    f(\alpha)/f(\beta)=f(\alpha-\beta)~.
\end{equation}
First, check the condition \eqref{g-condition}:
\begin{align}\begin{split}
    \frac{\phi(g+\alpha,g+\beta)}{\phi(\alpha,\beta)}
    &=\frac{\omega(\beta-\alpha,\alpha-\beta,-\alpha)}{\omega(\beta-\alpha,\alpha-\beta,-\alpha-g)}\\
    &=\theta_{\alpha,\beta}^g\theta_{\beta,\alpha}^g~,
\end{split}\end{align}
where we have used Eq. \eqref{thetatheta-ratio}. Now check Eq. \eqref{f-condition}:
\begin{align}\begin{split}
    \frac{\phi(\alpha,\beta)}{\phi(\beta,\alpha)}&=\frac{C(\alpha-\beta)}{C(\beta-\alpha)}\frac{\omega(\alpha-\beta,\beta-\alpha,-\beta)}{\omega(\beta-\alpha,\alpha-\beta,-\alpha)}\\
    &=\frac{C(\alpha-\beta)}{C(\beta-\alpha)}\omega(\alpha-\beta,\beta-\alpha,\alpha-\beta)~.
\end{split}\end{align}
For $\alpha-\beta=0$, this is trivial. For $\alpha-\beta=1$, this is
\begin{align}\begin{split}
    &=\frac{C(1)}{C(3)}\omega(1,3,1)=\frac{1}{f(3)}=f(1)~.
\end{split}\end{align}
For $\alpha-\beta=2$, this is
\begin{equation}
    \omega(2,2,2)=\omega(1,1,2)\omega(1,2,2)\omega(1,3,2)=f(2)~.
\end{equation}
For $\alpha-\beta=3$, this is
\begin{align}\begin{split}
    &=\frac{C(3)}{C(1)}\omega(3,1,3)=f(3)\omega(1,3,1)\omega(3,1,3)=f(3)~.
\end{split}\end{align}
Therefore, we conclude that the expression \eqref{phi-solution} is a solution to the conditions \eqref{g-condition} and \eqref{f-condition}. It would be useful to generalize this solution to other groups $G$, but we leave this for future work.

\subsubsection{Beyond square regions}\label{app:shape}

Now let us generalize the argument to arbitrary contractible and rotationally symmetric regions $\cD$. The boundary $\partial\cD$ can be partitioned into four similar intervals, each mapped into another by rotation. Denote the edges of the lattice that cross each of the four components by $e_\ell^\star$, $\star=1,2,3,4$, $\ell=1,\ldots,L$, labeled running clockwise. Each edge $e_\ell^\star$ is associated with two subsites in $\cD$ carrying degrees of freedom $i_{\ell,1}^\star,i_{\ell,2}^\star\in |G|$. Below we will use the superscript $\uparrow$ to denote that a term is present only when the edge $e_\ell^\star$ is ``up-facing'' (meaning it is vertically aligned with the site below it in $\cD$ and the site above it not in $\cD$); similarly with the superscripts $\rightarrow$, $\downarrow$, $\leftarrow$ for right-, down-, and left-facing edges. In the square region computation of \cref{sec:groupcohom}, the arrow superscripts were redundant with the component labels $\star$, but here they are distinct: each of the four intervals may contain edges facing in any direction.

Let $\alpha$ denote a configuration of labels $i_{s,r}$ satisfying the constraint $i_{s,3}=i_{s+v_1,4}=i_{s+v_1+v_2,1}=i_{s+v_2,2}$ that each plaquette has a GHZ state, and let $\alpha'$ be the configuration resulting from the action of the symmetry. Because of this constraint, both symmetries act on the wavefunction like operators on the boundary of $\cD$. Then the $G$ action becomes
\begin{align}\begin{split}
    U_g^\cD|\Psi_G\rangle
    &=\sum_\alpha\prod_{s\in D}\frac{\theta_{1,2}^{s,g}\theta_{2,3}^{s,g}}{\theta_{4,3}^{s,g}\theta_{1,4}^{s,g}}|\alpha'\rangle\\
    &=\sum_\alpha\prod_{\star=1}^4\prod_{\ell=1}^L\frac{\theta_{i_{\ell,1}^\star,i_{\ell,2}^\star}^g{}^{\uparrow\,+\,\rightarrow}}{\theta_{i_{\ell,2}^\star,i_{\ell,1}^\star}^g{}^{\downarrow\,+\,\leftarrow}}|\alpha'\rangle~,
\end{split}\end{align}
where again $\uparrow$ denotes $\delta(e_\ell^\star\text{ faces }\uparrow)\in\{0,1\}$, etc. In other words, each edge $e_\ell^\star$ in $\partial\cD$ contributes $\theta_{i_{\ell,1}^\star,i_{\ell,2}^\star}^g$ (if up- or right-facing) or $(\theta_{i_{\ell,2}^\star,i_{\ell,1}^\star}^g)^{-1}$ (if down- or left-facing). Meanwhile, by an argument like Eq. \eqref{f-arg}, the rotation action becomes
\begin{align}\begin{split}
    C_4^\cD|\Psi_G\rangle
    &=\sum_\alpha\prod_{s\in D}\frac{\phi(i_{s,4},i_{s,1})}{\phi(i_{s,2},i_{s,3})}|\alpha'\rangle\\
    &=\sum_\alpha\prod_{\star=1}^4\prod_{\ell=1}^L\phi(i_{\ell,1}^\star,i_{\ell,2}^\star)^{\leftarrow\,-\,\rightarrow}\\
    &\qquad\times\,f(i_{\ell,2}^\star)^{X-Y}|\alpha'\rangle~,
\end{split}\end{align}
which means each edge $e_\ell^\star$ contributes $\phi(i_{\ell,1}^\star,i_{\ell,2}^\star)$ (if left-facing), its inverse (if right-facing), $f(i_{\ell,1}^\star)$ or its inverse if the following conditions are satisfied
\begin{align}\begin{split}
    X&=\left\{\begin{array}{lr}1\qquad\,&\begin{array}{l}e_\ell^\star\text{ is }\downarrow\text{ and }e_{\ell+1}^\star\text{ is }\downarrow\\e_\ell^\star\text{ is }\downarrow\text{ and }e_{\ell+1}^\star\text{ is }\rightarrow\\e_\ell^\star\text{ is }\leftarrow\text{ and }e_{\ell+1}^\star\text{ is }\downarrow\\\,\end{array}\\0&\text{else}\end{array}\right.~,\\
    Y&=\left\{\begin{array}{lr}1\qquad\,&\begin{array}{l}e_\ell^\star\text{ is }\uparrow\text{ and }e_{\ell+1}^\star\text{ is }\uparrow\\e_\ell^\star\text{ is }\uparrow\text{ and }e_{\ell+1}^\star\text{ is }\leftarrow\\e_\ell^\star\text{ is }\rightarrow\text{ and }e_{\ell+1}^\star\text{ is }\uparrow\\\,\end{array}\\0&\text{else}\end{array}\right.~,
\end{split}\end{align}
where $e_{L+1}^\star$ means $e_1^{\star+1}$. Essentially, straight top and bottom edges and concave corners contribute. Let us denote the factors involving $f$ as $\cF$, i.e.
\begin{equation}
    \cF(\alpha):=\prod_{\star=1}^4\prod_{\ell=1}^Lf(i_{\ell,2}^\star)^{X-Y}~.
\end{equation}

Now put these symmetry actions together:
\begin{align}\begin{split}
    &I_{g,4}=\langle\Psi_G|U_g^\cD C_4^\cD|\Psi_G\rangle\\
    &=\sum_\alpha\langle\alpha|\alpha'\rangle\cF(\alpha)\prod_{\ell=1}^L\prod_{\star=1}^4\frac{\theta_{i_{\ell,1}^\star,i_{\ell,2}^\star}^g{}^{\leftarrow\,+\uparrow}}{\theta_{i_{\ell,2}^\star,i_{\ell,1}^\star}^g{}^{\rightarrow\,+\,\downarrow}}\phi(i_{\ell,1}^\star,i_{\ell,2}^\star)^{\leftarrow\,-\,\rightarrow}\\
    &=\sum_\alpha\cF(\alpha)\prod_{\ell=1}^L\prod_{\star=1}^4\delta(g\cdot i_{\ell,k}^\star=i_{\ell,k}^{\star+1})\\
    &\qquad\times\,\theta_{i_{\ell,1}^\star,i_{\ell,2}^\star}^g\cancel{\left(\frac{\phi(i_{\ell,1}^\star,i_{\ell,2}^\star)}{\phi(gi_{\ell,1}^\star,gi_{\ell,2}^\star)}\right)^{\rightarrow\,+\,\downarrow}\phi(i_{\ell,1}^\star,i_{\ell,2}^\star)^{\leftarrow\,-\,\rightarrow}}\\
    &=\pm\sum_{i_{\ell,1},i_{\ell_2}}\prod_{\ell=1}^L\prod_{j=1}^4\theta_{g^j\cdot i_{\ell,1},g^j\cdot i_{\ell,2}}^g\\
    &\qquad\times\,\delta(i_{\ell,2}=i_{\ell+1,1})\delta(i_{L,2}=g\cdot i_{1,1})\\
    &=\pm\sum_{i_{\ell,1}}\prod_{j=1}^4\omega(i_{L,1}^{-1}gi_{1,1},i_{1,1}^{-1}g^j,g)\\
    &\qquad\times\,\prod_{\ell=1}^{L-1}\omega(i_{\ell,1}^{-1}i_{\ell+1,1},i_{\ell+1,1}^{-1}g^j,g)\\
    &=\pm\sum_{i_{\ell,1}}\prod_{j=1}^4\omega(g,g^j,g)\cancel{\frac{\omega(i_{L,1}^{-1},g^j,g)}{\omega(i_{1,1}^{-1},g^j,g)}\prod_{\ell=1}^{L-1}\frac{\omega(i_{\ell,1}^{-1},g^j,g)}{\omega(i_{\ell+1,1}^{-1},g^j,g)}}\\
    &\sim\pm\cZ_\omega(L(4;1)_g)~.
\end{split}\end{align}
In the above, we have used the fact that $\langle \alpha|\alpha'\rangle$ imposes the constraint $g\cdot i_{\ell,k}^\star=i_{\ell,k}^{\star+1}$ in order to translate the computation to one of the four boundary components, e.g. $\star=1$ (thus we drop the $\star$ superscript). In the penultimate line, we used the cocycle relation \eqref{cocycle-product} with $h=g^{-1}i_{L,1}$, $k=i_{1,1}$ and with $h=i_{\ell,1}$, $k=i_{\ell+1,1}$. The $\pm$ sign comes from the $\cF(\alpha)$ factor. Under the constraints on $i$, it becomes
\begin{align}\begin{split}
    &\prod_{\star=1}^4\prod_{\ell=1}^Lf(i_{\ell,2}^\star)^{X-Y}=\prod_{\ell=1}^L\prod_{\star=1}^4\left(\frac{f(g^2\cdot i_{\ell,2}^\star)}{f(i_{\ell,2}^\star)}\right)^Y\\
    &\qquad\qquad\quad=\prod_{\star=1}^4\prod_{\ell=1}^L\left(\frac{\phi(g^2\cdot i_{\ell,2}^\star,i_{\ell,2}^\star)}{\phi(i_{\ell,2}^\star,g^2\cdot i_{\ell,2}^\star)}\right)^Y\\
    &\qquad\qquad\quad=\prod_{\star=1}^4\prod_{\ell=1}^L\left(\theta_{i_{\ell,2}^\star,g^2\cdot i_{\ell,2}^\star}^{g^2}\theta_{g^2\cdot i_{\ell,2}^\star,i_{\ell,2}^\star}^{g^2}\right)^Y\\
    &\qquad\qquad\quad=\prod_{\star=1}^4\prod_{\ell=1}^L\left(\frac{\omega(g^2,g^2,(i_{\ell,2}^{\star})^{-1})}{\omega(g^2,g^2,(i_{\ell,2}^{\star})^{-1}g^2)}\right)^Y\\
    &\qquad\qquad\quad=\prod_{\star=1}^4\prod_{\ell=1}^L\left(\frac{\omega(g^2,g^2,(i_{\ell,2}^{\star})^{-1})}{\omega(g^2,g^2,g^2(i_{\ell,2}^{\star})^{-1})}\right)^Y\\
    &\qquad\qquad\quad=\omega(g^2,g^2,g^2)^{\#(Y)}\\
    &\qquad\qquad\quad=\pm 1~.
\end{split}\end{align}
Here, we used the $g^2$ symmetry to rotate all $X=1$ edge to $Y=1$ edges, then used Eq. \eqref{f-condition} followed by Eq. \eqref{g-condition} and the relation \eqref{thetatheta-ratio}. As in \cref{sec:groupcohom}, we must assume that $g^2$ is central in $G$. We then used the cocycle condition for $(g^2,g^2,g^2,(i_{\ell,2}^\star)^{-1})$ and finally that $\omega(g^2,g^2,g^2)=\cZ_\omega(L(2;1)_{g^2})\in\{\pm 1\}$.

To determine the sign for a particular region $\cD$, it remains to count the number $\#(Y)$ of edges in $\partial\cD$ that satisfy the condition $Y$. Consider deforming the region by adding or subtracting a single site from each of the four components $\star$ such that the region remains rotationally symmetric. By checking a few cases, one can verify that this deformation can only ever change $\#(Y)$ by an even number. This means that the sign $\cF(\alpha)$ is shape-independent. In particular, the results for square regions apply: the sign is $+1$ for vertex-centered regions and $\cZ_\omega(L(2;1)_{g^2})$ for plaquette-centered regions.

\subsubsection{Modified group cohomology models}\label{app:modified}

Now we consider an alternative construction to the usual group cohomology models. These models allow us to compute the partial symmetry order using the usual rotational symmetry that acts like
\begin{equation}\label{plain-rotation}
    \hat C_4^s|i_{s,1},i_{s,2},i_{s,3},i_{s,4}\rangle=|i_{s',2},i_{s',3},i_{s',4},i_{s',1}\rangle~.
\end{equation}
This operation does not commute with the usual on-site $G$ symmetry \eqref{G-sym}. To remedy this issue, we modify the $G$ symmetry action to be
\begin{align}\begin{split}\label{newsym}
    \tilde U_g&|i_{s,1},i_{s,2},i_{s,3},i_{s,4}\rangle
    \\&\quad=\theta_{1,2}^{s,g}\theta_{2,3}^{s,g}\theta_{3,4}^{s,g}\theta_{4,1}^{s,g}|gi_{s,1},gi_{s,2},gi_{s,3},gi_{s,4}\rangle~
\end{split}\end{align}
This representation is captured by the state sum of the lens space partition function depicted in \cref{fig:statesum}, where each tetrahedron yields a factor of $\omega$. Note the rotational symmetry of this diagram, in contrast with \cref{fig:groupcohom-action} for the $G$ action in the unmodified models \cite{Chen2013}. From this diagram, one can see that $\tilde U_g$ forms a linear representation of $G$, using an argument similar to that in appendix G of Ref. \cite{Chen2013}.

Finally, the wavefunction must be modified to one that is invariant under the new symmetry \eqref{newsym}. To achieve this, add a weight to each configuration $\{\alpha_p\}$ of plaquette labels:
\begin{equation}\label{mod-wavefunction}
    |\tilde\Psi_G\rangle=\sum_{\alpha_p\in G}\Phi(\{\alpha_p\})\bigotimes_p|\alpha_p,\alpha_p,\alpha_p,\alpha_p\rangle~.
\end{equation}
Suppose the weight $\Phi$ has the form
\begin{equation}
    \Phi(\{\alpha_p\})=\prod_s\phi(i_{s,1},i_{s,2})\phi(i_{s,2},i_{s,3})~,
\end{equation}
where the subsite labels $i_{s,r}$ come from the $\alpha_p$. The asymmetric choice of subscripts is made so that each edge (horizontal or vertical) gets one factor of $\phi$ (from the vertex below or to the left, respectively). Under the action \eqref{newsym} of the $G$ symmetry, each edge picks up a $\theta$ from each of its two adjacent vertices, for a total of $\theta_{ij}^g\theta_{ji}^g$. For the state to be invariant, this factor must compensate for the change in weight from $\phi(i,j)$ to $\phi(gi,gj)$; this condition is precisely the condition \eqref{g-condition} we identified previously.

Despite the asymmetry of this ansatz, if we can find a solution $\phi$ satisfying our other condition \eqref{f-condition}, the modified wavefunction \eqref{mod-wavefunction} is nevertheless invariant under the rotation symmetry \eqref{plain-rotation}. To see this, note that under a quarter rotation, the horizontal edge factors $\phi(i_{s,2},i_{s,3})$ map to the vertical edge factors $\phi(i_{s',1},i_{s',2})$. On the other hand, $\phi(i_{s,1},i_{s,2})$ maps to $\phi(i_{s',4},i_{s',1})=\phi(i_{s'-v_1,3},i_{s'-v_2,2})$, with the $2$ and $3$ subscripts reversed. Thus, rotation invariance requires that the product of ratios $\phi(i,j)/\phi(j,i)$ over the whole lattice is $1$. By an argument like Eq. \eqref{f-arg}, this is ensured by the condition \eqref{f-condition}. We remark that because of the asymmetry of the ansatz, it is unclear how to extend the modified group cohomology model to other lattices or to lattice defects.

Now we can compute the order parameter. Consider $4$-fold rotations around a single vertex. The $\phi$ weights on the bra and ket cancel, leaving
\begin{align}\begin{split}
    I_{g,4}^v
    &=\langle\Psi_G|\tilde U_gC_4^s|\Psi_G\rangle\\
    &=\sum_{i_1,i_2,i_3,i_4}\langle i_1,i_2,i_3,i_4|\tilde U_gC_4^s|i_1,i_2,i_3,i_4\rangle\\
    &=\sum_{i_1,i_2,i_3,i_4}\theta_{1,2}^g\theta_{2,3}^g\theta_{3,4}^g\theta_{4,1}^g\prod_{k=1}^4\delta(g\cdot i_k=i_{[k+1]})\\
    &\sim\cZ_\omega(L(4;1)_g)~.
\end{split}\end{align}
Partial symmetry order parameters for order $2$ rotations and other regions may be computed similarly.

\section{Disentangling $k_1$, $k_2$ and $k_3$ for various rotation centers}\label{app:determine}

Consider a $\Z_n$ on-site symmetry together with the spatial rotation symmetries of a square or honeycomb lattice. In this section we study the extent to which the on-site SPT invariant $k_1$ defined in \cref{sec:cryst} can be recovered through partial symmetry order parameters alone. We will work with a single origin of rotations. For best results, use an origin with a maximal degree of symmetry (e.g. a face of the honeycomb lattice); since this degree is divisible by the degrees at other origins, partial symmetries at this origin capture as much information about $k_1$ as do partial symmetries at any origin.

Recall that the order parameter for an $M$-fold rotation rotation together with an on-site rotation by $g=q \mod n$ takes the form
\begin{align}\begin{split}\label{beta}
    I_{g,M} &:= e^{i \frac{2\pi}{M}\beta_{\alpha,M}}~,\\
    \beta_{\alpha,M} &= \alpha^2 k_1 + \alpha k_2 + k_3 \mod M~.
\end{split}\end{align}
where $\alpha = q M/n$. By choosing any on-site symmetry $q$ such that $qM/n$ is an integer, the parameter $\alpha$ can obtain integer values that are multiples of $M/(n,M)$; mod $M$, there are $(n,M)$ such values of $\alpha$. We can also consider different values of the rotation parameter $M$ that divide the order of the total rotational symmetry around the rotation center. We thus obtain a system of linear congruence relations which we can attempt to solve for $k_1$.

\subsubsection{$2$-fold rotational symmetry}

In this case, we can only perform $M=2$ rotations. For $n$ even, there are $(n,M)=2$ values of $\alpha$ mod $M$, given by $\alpha=0,1$. From the expression \eqref{beta}, we obtain the two relations
\begin{align}\begin{split}
    \beta_{0,2} &= k_3 \mod 2~,\\
    \beta_{1,2} &= k_1 + k_2 + k_3 \mod 2~,
\end{split}\end{align}
which allows us determine $k_3\in\ZZ_2$ exactly as well as $k_1+k_2$ mod $2$. The on-site invariant $k_1$ cannot be independently determined. On the other hand, for $n$ odd, we only have the $\alpha=0$ relation $\beta_{0,2}=k_3$, as $\beta_{1,2}$ is undefined. This determines $k_3$; meanwhile, $k_2\in\ZZ_{(n,M)}$ does not exist, and $k_1$ is undetermined.

In light of this analysis, we can conclude that the CZX state has a nontrivial mixed SPT invariant $k_2$ for edge-centered rotations. Indeed, we computed in \cref{sec:CZX} that the CZX state has $I_{{\bf g},2}^e=I_{{\bf e},2}^e=+1$, i.e. $k_1+k_2^e=0 \mod 2$. Since we know $k_1=1$ (from using other rotation centers), this implies $k_2^e=1$.

\subsubsection{$3$-fold rotational symmetry}

Again, we only have rotations of a single order $M=3$. As always, if $n$ and $M$ are relatively prime (here, if $n$ is not a multiple of $3$), only the $\alpha=0$ relation holds, so we can determine $k_3$ but not $k_1$ (and $k_2$ does not exist). If $n$ is a multiple of $3$, we have three relations, corresponding to $\alpha=0,1,2$:
\begin{align}\begin{split}
    \beta_{0,3} &= k_3 \mod 3~, \\
    \beta_{1,3} &= k_1 + k_2 + k_3 \mod 3~,\\
    \beta_{2,3} &= k_1 + 2 k_2 + k_3 \mod 3~.\\
\end{split}\end{align}
These allow us to solve for $k_1, k_2$ mod $3$ as
\begin{align}\begin{split}
    k_1 &= 2\beta_{0,3} + 2\beta_{1,3} + 2\beta_{2,3} \mod 3~,\\
    k_2 &= 2\beta_{1,3} + \beta_{2,3} \mod 3~.
\end{split}\end{align}
In other words, for $\ZZ_n=\{{\bf e},{\bf g},{\bf g}^2,\ldots,{\bf g}^{n-1}\}$ the partial symmetries one needs to measure are
\begin{align}\begin{split}
    e^{\frac{2\pi i}{3}k_1}&=(I_{{\bf e},3}I_{{\bf g}^{n/3},3}I_{{\bf g}^{2n/3},3})^2~,\\
    e^{\frac{2\pi i}{3}k_2}&=(I_{{\bf g}^{n/3},3})^2I_{{\bf g}^{2n/3},3}~,\\
    e^{\frac{2\pi i}{3}k_3}&=I_{{\bf e},3}~.
\end{split}\end{align}
Note that for $n=6,9,\ldots$, the invariant $k_1\in\ZZ_n$ is not determined absolutely, only mod $3$.

\subsubsection{$4$-fold rotational symmetry}

In this setting, we have both $M=4$ and $M=2$ partial rotations at our disposal.

We divide the analysis into three cases. 

\textbf{Case 1}: $n$ is odd. Since $n$ is relatively prime to both $M=4$ and $M=2$, there is only the $\alpha=0$ condition, so we learn nothing about $k_1$.

\textbf{Case 2}: $n \equiv 2 \mod 4$. With order $M=4$ rotations, we have the $\alpha = 0,2$ relations
\begin{align}\begin{split}
    \beta_{0,4} &= k_3 \mod 4~, \\
    \beta_{2,4} &= 2 k_2 + k_3 \mod 4~.
\end{split}\end{align}
We can also use order $M=2$ rotations about the same origin. Now
$\alpha = 0,1$, so we have relations
\begin{align}\begin{split}
    \beta_{0,2} &= k_3 \mod 2~,\\
    \beta_{1,2} &= k_1 + k_2 + k_3 \mod 2~.
\end{split}\end{align}
From these relations, we solve for $k_1$ and $k_2$ mod $2$:
\begin{align}\begin{split}
    k_1 &= \beta_{1,2} + \tfrac{1}{2}(\beta_{0,4}+\beta_{2,4}) \mod 2~,\\
    k_2 &= \tfrac{1}{2}(\beta_{0,4}-\beta_{2,4}) \mod 2~.
\end{split}\end{align}
In other words, for $\ZZ_n=\{{\bf e},{\bf g},{\bf g}^2,\ldots,{\bf g}^{n-1}\}$ the partial symmetries one needs to measure are
\begin{align}\begin{split}
    e^{\frac{2\pi i}{2}k_1}&=I_{{\bf e},4}I_{{\bf g}^{n/2},4}I_{{\bf g}^{n/2},2}~,\\
    e^{\frac{2\pi i}{2}k_2}&=I_{{\bf e},4}/I_{{\bf g}^{n/2},4}~,\\
    e^{\frac{2\pi i}{4}k_3}&=I_{{\bf e},4}~.
\end{split}\end{align}
This result generalizes Eq.~\eqref{eq:k1-invt} from $n=2$. Note that for $n=6,10,\ldots$, the invariant $k_1\in\ZZ_n$ is not determined absolutely, only mod $2$.

\textbf{Case 3}: $n \equiv 0 \mod 4.$ With order $M=4$ rotations, we have the $\alpha = 0,1,2,3$ relations
\begin{align}\begin{split}
    \beta_{0,4} &= k_3 \mod 4~,\\
    \beta_{1,4} &= k_1 + k_2 + k_3 \mod 4~,\\
    \beta_{2,4} &= 2 k_2 + k_3 \mod 4~,\\
    \beta_{3,4} &= k_1 + 3 k_2 + k_3 \mod 4~.
\end{split}\end{align}
Order $M=2$ rotations have $\alpha=0,1$ and give
\begin{align}\begin{split}
    \beta_{0,2} &= k_3 \mod 2~,\\
    \beta_{1,2} &= k_1 + k_2 + k_3 \mod 2~,
\end{split}\end{align}
so, unlike in the previous case, order $2$ rotations provide no new information over order $4$ rotations. From these relations, we can solve for the quantities
\begin{align}\begin{split}
    k_1 + k_2 &= \beta_{1,4} - \beta_{0,4} \mod 4~,\\
    2k_1 &= 2\beta_{0,4} + \beta_{1,4} + \beta_{3,4} \mod 4~,\\
    2k_2 &= \beta_{2,4} - \beta_{0,4} \mod 4~,
\end{split}\end{align}
which determines $k_1$ and $k_2$ mod $2$, but there is no way to determine $k_1\in\ZZ_n$ and $k_2\in\ZZ_4$ absolutely. The measurements one must make are the following:
\begin{align}\begin{split}
    e^{\frac{2\pi i}{4}(k_1+k_2)}&=I_{{\bf g}^{n/4},4}/I_{{\bf e},4}~,\\
    e^{\frac{2\pi i}{2}k_1}&=(I_{{\bf e},4})^2I_{{\bf g}^{n/4},4}I_{{\bf g}^{3n/4},4}~,\\
    e^{\frac{2\pi i}{2}k_2}&=I_{{\bf g}^{n/2},4}/I_{{\bf e},4}~,\\
    e^{\frac{2\pi i}{4}k_3}&=I_{{\bf g},4}~.
\end{split}\end{align}

\subsubsection{$6$-fold rotational symmetry}
Here it is sufficient to combine the equations for $M=2,3$. In particular, if $n$ is even, we get $k_1 + k_2$ mod $2$, while if 3 divides $n$, we get $k_1$ mod $3$. 



\bibliography{ref.bib}


\end{document}